\documentclass[%
 aip,
 amsmath,amssymb,
 reprint,%
]{revtex4-1}

\usepackage{graphicx}
\usepackage{dcolumn}
\usepackage{bm}

\usepackage[utf8]{inputenc}
\usepackage[T1]{fontenc}
\usepackage{mathptmx}
\usepackage{etoolbox}
\usepackage{graphicx}
\usepackage{dcolumn}
\usepackage{bm}
\usepackage{xcolor}
\usepackage{physics}
\usepackage[utf8]{inputenc}
\usepackage[T1]{fontenc}
\usepackage{mathptmx}
\usepackage{etoolbox}
\usepackage{graphicx}  
\usepackage{subfig}
\usepackage{multirow}
\usepackage{xcolor}
\usepackage{fancyhdr}
\usepackage{parskip}
\usepackage[T1]{fontenc}
\usepackage{dcolumn}   
\usepackage{bm}        
\usepackage{amsfonts}  
\usepackage{amsmath}   
\usepackage{amssymb}   
\usepackage{tikz,xcolor,hyperref}
\usepackage{hyperref}
\hypersetup{
    colorlinks=true,
    linkcolor=blue,
    filecolor=magenta,      
    urlcolor=cyan,
    pdftitle={Quantum metrology},
    pdfpagemode=FullScreen,
    citecolor=magenta,
    } 

\makeatletter
\def\@email#1#2{%
 \endgroup
 \patchcmd{\titleblock@produce}
  {\frontmatter@RRAPformat}
  {\frontmatter@RRAPformat{\produce@RRAP{*#1\href{mailto:#2}{#2}}}\frontmatter@RRAPformat}
  {}{}
}%
\makeatother

\definecolor{lime}{HTML}{A6CE39}
\DeclareRobustCommand{\orcidicon}{
	\begin{tikzpicture}
	\draw[lime, fill=lime] (0,0) 
	circle [radius=0.16] 
	node[white] {{\fontfamily{qag}\selectfont \tiny ID}};
	\draw[white, fill=white] (-0.0625,0.095) 
	circle [radius=0.007];
	\end{tikzpicture}
	\hspace{-2mm}
}

\foreach \x in {A, ..., Z}{%
	\expandafter\xdef\csname orcid\x\endcsname{\noexpand\href{https://orcid.org/\csname orcidauthor\x\endcsname}{\noexpand\orcidicon}}
}

\begin{document}

\preprint{AIP/123-QED}

\title[Quantum metrology in a lossless Mach-Zehnder interferometer using entangled photon inputs]{Quantum metrology in a lossless Mach-Zehnder interferometer using entangled photon inputs for a sequence of non-adaptive and adaptive measurements}
\author{Shreyas S\orcidA{}}
\email{ssadugol@tulane.edu}
\author{Lev Kaplan}%
 \email{lkaplan@tulane.edu}
\affiliation{ 
\it Department of Physics and Engineering Physics,
Tulane University, New Orleans, Louisiana 70118, USA
}%



\date{\today}

\begin{abstract}
Using multi-photon entangled input states, we estimate the phase uncertainty in a noiseless Mach-Zehnder interferometer (MZI) using photon-counting detection. We assume a flat prior uncertainty and use Bayesian inference to construct a posterior uncertainty.
By minimizing the posterior variance to get the optimal input states, we first devise an estimation and measurement strategy that yields the lowest phase uncertainty for a single measurement. N00N and Gaussian states are determined to be optimal in certain regimes.
We then generalize to a sequence of repeated measurements, using non-adaptive and fully adaptive measurements. N00N and Gaussian input states are close to optimal in these cases as well, and optimal analytical formulae are developed. Using these formulae as inputs, a general scaling formula is obtained, which shows how many shots it would take on average to reduce phase uncertainty to a target level. Finally, these theoretical results are compared with a Monte Carlo simulation using frequentist inference. In both methods of inference, the local non-adaptive method is shown to be the most effective practical method to reduce phase uncertainty. 
\end{abstract}

\maketitle

\section{\label{sec:Introduction}Introduction}

Quantum metrology is the science of measurements using quantum systems. The goal is to achieve the ultimate fundamental bounds on estimation precision of unknown parameters, using non-classical input states or probes. Quantum enhanced precision measurements have notable applications in biological systems~\cite{app:biology}, gravitational wave detection~\cite{app:grav}, atomic clocks~\cite{app:atomic_clock_1,app:atomic_clock_2}, Hamiltonian estimation~\cite{Hamiltonian_Learning}, sensing~\cite{app:sensing} and imaging~\cite{app:imaging_1, app:imaging_2}. 

Photons are popular quantum systems due to their generation, manipulation, and detection properties~\cite{Polino_et_al}. Since many physical problems can be mapped into phase estimation processes \cite{Polino_et_al}, it is useful to study them using a lossless Mach-Zehnder or other mathematically equivalent SU(2) interferometer~\cite{SU(2)_interferometry, rosetta, Polino_et_al, Lee_et_al}. Since there is no Hermitian quantum phase operator \cite{phase_operator}, the phase $\phi$ needs to estimated and cannot be directly measured. Its estimation is done by decrypting the measurement statistics of the photon number operator, which is Hermitian, by photon counting at the output.

Different approaches and optimal estimators exist: Maximum likelihood estimator (MLE), method of moments, and Bayesian estimator are a few of the notable ones. While the first two are based on an objective, frequentist interpretation of probability, the Bayesian approach interprets probability subjectively~\cite{Polino_et_al, Pezze}. The Bayesian approach requires us to construct a posterior probability distribution given a prior distribution after evidence is gathered.

The lowest achievable phase estimation error $\Delta\phi$ for a single measurement using a classical $N$-photon state  scales in accordance with the standard quantum limit (SQL): $\Delta\phi\sim\frac{1}{\sqrt{N}}$. This limit is also known as the shot noise limit (SNL). Multiple papers have shown that the Heisenberg limit (HL):  $\Delta\phi\sim\frac{1}{N}$ is achievable in some circumstances using non-classical or entangled states \cite{entanglement_0,HL_Limit_0,HL_Limit_1,rosetta}. For example, Ref.~\cite{adaptive} showed that HL scaling can be achieved in a canonical measurement framework (but not using photocounting unless an adaptive scheme is implemented where the interferometer is adjusted after each photon is measured for an $N$-photon entangled state).
While it is necessary to use entangled input states to increase the phase estimation precision beyond the SQL, entanglement is useless at the measurement stage~\cite{metrology-Giovannetti}. Also, not all entangled states provide sub-SNL precision~\cite{useful_entanglement-Pezze,entanglement_1,entanglement_2}. The condition $F\geq N$, where $F$ is the Fisher information, is sufficient for entanglement and necessary and sufficient for a state to be useful to achieve a sub-SNL sensitivity~\cite{Pezze,useful_entanglement-Pezze}.

Bayesian inference also allows the development of adaptive measurement strategies \cite{adaptive}. The HL lower bound can be further lowered by making multiple independent measurements or shots $\nu$. By doing so we expect, from the central limit theorem, that we obtain scaling $\Delta\phi \sim \frac{1}{N\sqrt{\nu}}$, to be compared with $\Delta\phi \sim \frac{1}{N}$ for a single shot. As a consequence of the Laplace-Bernstein-von Mises theorem, asymptotically in $\nu$, the posterior variance becomes normally distributed and centered at the true value of the parameter, with a variance inversely proportional to the Fisher Information~\cite{Pezze, Kay}. 

For these reasons, in this work we input entangled photons into a lossless MZI and make measurements by photon counting. We then use the measurement statistics to estimate the true value of the phase, using a sequence of adaptive and non-adaptive measurements, while reducing the posterior variance. Extensive work has been done on lossy MZI-like interferometers, using special GPCS states~\cite{N00N_states_optimal, lossy_optimization_2}. These studies have applications in the optimization of linear optical systems\cite{Lee_et_al, LOQC_1, LOQC_2, LOQC_3}. 

\section{Single Measurement}
\subsection{Mathematical Framework}

We seek to estimate the phase shift $\phi$ between the two interferometer arms by measuring the number of photons at the detectors D1 and D2 (see Fig.~\ref{fig:MZI}). Since the parameter we are attempting to estimate is nontrivially related to the measurement outcome $m$, we follow the Bayesian framework. In this framework, we assume that $\phi$ is a random variable with a flat prior probability density function, PDF $p(\phi)$, which represents the complete lack of knowledge of $\phi$ prior to starting the experiment. Without loss of generality, we can set the mean of $p(\phi)$ to 0.

 \begin{figure}[h]
    \centering
    \includegraphics[width=\columnwidth]{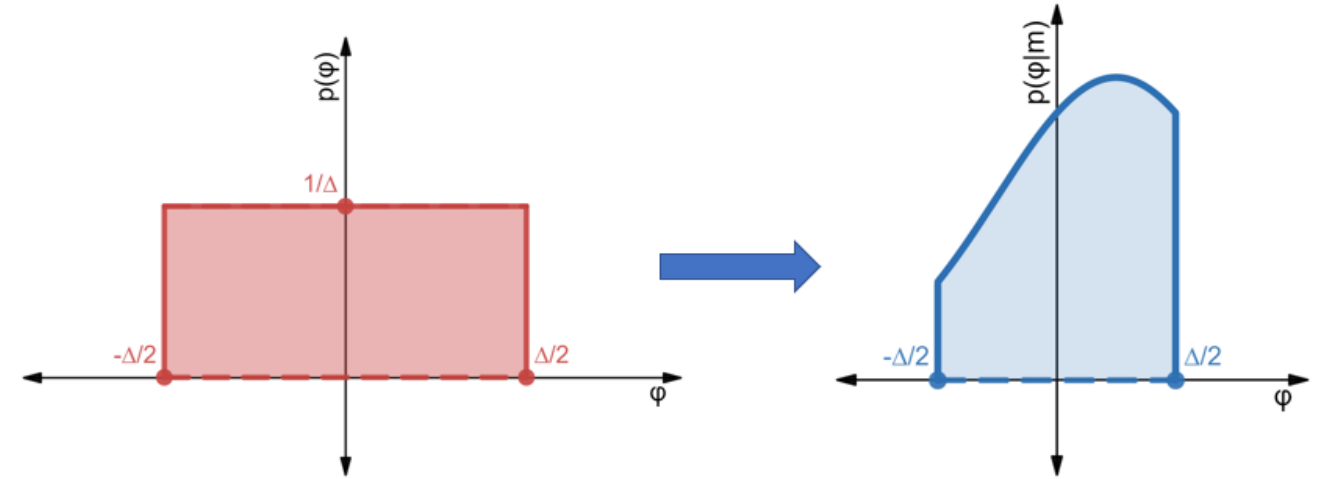}
    \caption{Flat prior PDF $p(\phi)$ (left) and a hypothetical posterior PDF $p(\phi|m)$ for a particular measurement outcome $m$ (right).}
    \label{fig:comparison of prior and posterior PDFs}
\end{figure}

Our choice of estimator $\tilde{\phi}$ needs to help us determine $\phi$ to the highest degree of accuracy. In other words, $\tilde{\phi}$ should minimize the Bayesian mean square error (BMSE) of the true value of $\phi$, where $ BMSE(\tilde{\phi}) = E[(\tilde{\phi}-\phi)^2]$. It can be shown that $\tilde{\phi}_m= E[\phi|m]$~\cite{Kay}. This is the mean of the posterior PDF and is usually called minimum mean squared error (MMSE). Given the prior PDFs, $p(\phi)$ and $p(m|\phi)$, we can use Bayes' Theorem to give us the posterior PDF $p(\phi|m)$, which for a single measurement outcome $m$ is given by~\eqref{eq: Bayes Theorem single shot}

\begin{equation}
\label{eq: Bayes Theorem single shot}
p(\phi|m)=\frac{p(\phi)p(m|\phi)}{p(m)}    \,.
\end{equation}

For a single shot ($\nu=1$) and with appropriate integration limits, our estimator is given by:
\begin{equation}
\label{eq:phitilde}
    \tilde{\phi}_m= E[\phi|m] = \int \phi\,p(\phi|m)\;d\phi = \frac{\int_{-\Delta/2}^{\Delta/2} \phi\,p(\phi) p(m|\phi) \,d\phi\ }{p(m)}
\end{equation}

\noindent where $p(m)$ is:
\begin{equation}
    p(m) = \int_{-\Delta/2}^{\Delta/2} p(\phi)p(m|\phi) \,d\phi\ \label{Pm} \,.
\end{equation}

\noindent Similarly, the BMSE, also called the posterior variance (summing over the discrete outcomes $m$ after a single shot), takes the form:
\begin{equation}
\label{eq:BMSE2}
    BMSE(\tilde{\phi}) = \sum_m\int_{-\Delta/2}^{\Delta/2} (\phi - \tilde{\phi})^2 p(\phi) p(m|\phi) \, d\phi
\end{equation}

Eqs.~\eqref{eq:phitilde}, \eqref{Pm}, and \eqref{eq:BMSE2} can be explicitly evaluated once the conditional PDF $p(m|\phi)$ is determined. This conditional PDF is obtained using the physical framework as discussed in Sec.~\ref{Physical framework}.

Note that the optimal estimation strategy within the Bayesian approach depends explicitly on the prior PDF assumed and it is very important to choose an appropriate prior PDF. If $p(\phi)$ is sensitive to small changes in $\phi$ and robust to changes in $m$,  $p(\phi)p(m|\phi)\approx p(\phi)$, then the $BMSE(\tilde{\phi})$ will be dominated by the prior PDF and the gathered data will have limited effect on the estimation process~\cite{metrology_review}.

\subsection{Physical Framework}\label{Physical framework}

\begin{figure}[h]
    \centering
    \includegraphics[width=\columnwidth]{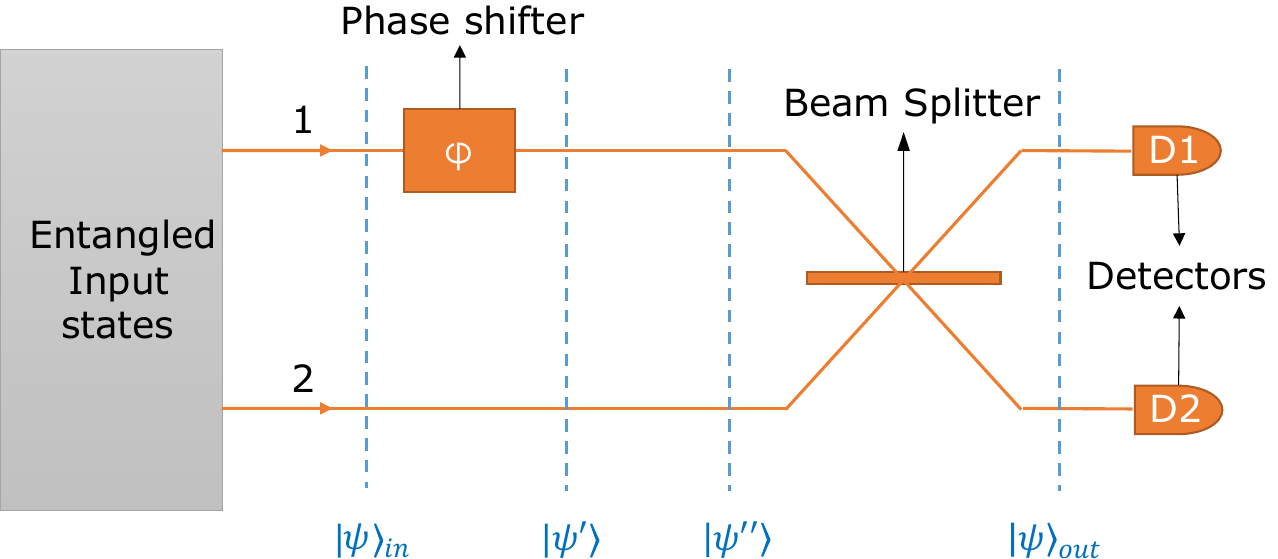}
    \caption{Schematics of a typical lossless MZI. The second detector is not required.}
    \label{fig:MZI}
\end{figure}

\noindent In Fock notation, assuming no loss, an arbitrary $N$-photon pure-state input is given by:

\begin{multline}
    \ket{\psi}_{in}=\sum_{k=0}^N c_k\ket{N-k,k} =
    \sum_{k=0}^N c_k \frac{(\hat{a}_1^\dagger)^{N-k}(\hat{a}_2^\dagger)^k}{\sqrt{(N-k)!k!}}\ket{0,0}\,,
    \label{inputstate}
\end{multline}
\noindent where $\ket{N-k,k}$ represents $N-k$ photons in arm 1 and $k$ photons in arm 2. $\hat{a}_i^\dagger$ is the photon creation operator in arm $i$, and  $c_k$ is the input state coefficient. In Fig.~\ref{fig:MZI}, $\ket{\psi}_{in}$ is the output of the gray box.

Next, $\ket{\psi}_{in}$ passes through the phase shifter $e^{i\phi\hat{n}_1}$, where $\hat{n}_1 = (N-k)$ is the number operator on arm 1, and evolves to $\ket{\psi'}$:

\begin{equation}
    \ket{\psi'} = \sum_{k=0}^N c_k \frac{e^{i\phi(N-k)}(\hat{a}_1^\dagger)^{N-k}(\hat{a}_2^\dagger)^k}{\sqrt{(N-k)!k!}}\ket{0,0}\,.
\end{equation}

For a lossy interferometer, photon loss is typically modeled by fictitious beam splitters that remove photons from the system, and $\ket{\psi'} \neq \ket{\psi''}$~\cite{Quantum_theory_of_light,Lee_et_al,LK_Preprint}. However, in the lossless case, which is a primary assumption of this paper,  $\ket{\psi'} = \ket{\psi''}$.
 
 The action of the beam splitter on $\ket{\psi''}$ is a unitary transformation on the creation operators for arms 1 and 2:

\begin{equation}
\label{eq:BS transform}
    \begin{pmatrix}
        \hat{a}_{1,out}^\dagger\\
        \hat{a}_{2,out}^\dagger
    \end{pmatrix}
    =
    U \begin{pmatrix}
        \hat{a}_{1,in}^\dagger\\
        \hat{a}_{2,in}^\dagger
    \end{pmatrix} \,,
\end{equation}
where the ``in" and ``out" subscripts label creation operators before and after the beam splitter, respectively, and the 2x2 matrix $U$ is
  \begin{equation}
  \label{U3}
         U=
    \begin{pmatrix}
        \cos(\gamma) & \sin(\gamma)\\
        -\sin(\gamma) & \cos(\gamma)\\
    \end{pmatrix}  \,.
 \end{equation} 
Therefore,
 \begin{multline}
         \ket{\psi'''} = \sum_{k=0}^N c_k \frac{e^{i\phi(N-k)}}{\sqrt{(N-k)!k!}}\\ \times [(\cos{\gamma}\hat{a}_1^\dagger + \sin\gamma\hat{a}_2^\dagger)]^{N-k} [(-\sin{\gamma}\hat{a}_1^\dagger + \cos\gamma\hat{a}_2^\dagger)]^{k}\ket{0,0} \,.
 \end{multline}
 
Without loss of generality, the beam splitter is taken to be a 50:50 beam splitter, i.e., $\gamma=\pi/4$. (If instead we let $\gamma$ be a free parameter, then optimizing the BMSE yields $\gamma=\pi/4$.) Intuitively, a 50:50 beam splitter provides maximal interference between the two MZI arms, and thus maximal information about the unknown phase $\phi$.

 Finally the photon detectors D1 and D2 make projective measurements on the state. These measurements are formally described by an observable $M$
 with spectral decomposition $M=\sum_m m P_m$ on the state space of $\ket{\psi'''}$, where $P_m$ is a projection operator of a particular outcome $m$. The probability that outcome $m$ occurs for a given phase difference $\phi$ is given by
 \begin{equation}
 \label{eq: pmphi}
     p(m|\phi) = \bra{\psi'''}P_m\ket{\psi'''} \,.
 \end{equation}
 
Given the expression \eqref{eq: pmphi} for $p(m|\phi)$, we see from \eqref{eq:BMSE2} that the BMSE depends only on the complex input state coefficients $c_k$,which can be expressed in polar form, $c_k = r_k e^{i\theta_k}$, where $r_k,\theta_k \in \mathbb{R}$.
\begin{figure}[h]
    \includegraphics[width=\columnwidth]{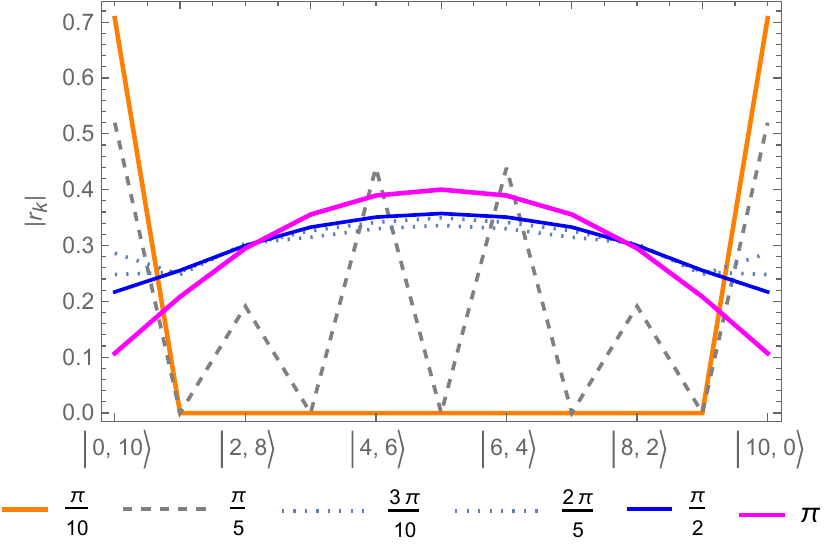}
    \caption{Optimal single-shot input states for $N=10$ photons and prior uncertainties $\Delta=\pi,\frac{\pi}{2}, \frac{\pi}{3}, \frac{3\pi}{10}, \frac{\pi}{5},\frac{\pi}{10}$. \textcolor{orange}{Orange line} corresponds to a N00N-like state, and \textcolor{magenta}{magenta} and \textcolor{blue}{blue lines} correspond to Gaussian or Quasi-Gaussian states. Dotted and dashed lines correspond to intermediate states.}
    \label{fig:Optimal single shot input states}
\end{figure}

We are then able to find the optimal BMSE by minimizing \eqref{eq:BMSE2} w.r.t. $r_k,\theta_k$. Fig.~\ref{fig:Optimal single shot input states} shows the optimal input amplitudes $|r_k|$ for 10 photons for various prior uncertainties $\Delta$. These optimal input state amplitudes can be categorized under 3 regimes: N00N, Intermediate, and quasi-Gaussian/Gaussian. These regimes depend on the photon number $N$ and the prior phase uncertainty $\Delta$. We restrict our prior phase uncertainty to $0\leq\Delta<\pi$ to eliminate ambiguities of the type $p(m|\phi) = p(m|\phi+\pi)$ when determining the estimator $\tilde{\phi}$.\\

The N00N and Gaussian states have analytical forms given by \eqref{N00N} and \eqref{Gaussian state}, where $s=\pm 1$ and $\rho$ is a function of $N$ and $\Delta$:
\begin{equation}
    \label{N00N}
    \ket{\psi}_{N00N} = \frac{\ket{N,0}+e^{\pm i\pi/2}\ket{0,N}}{\sqrt{2}}
\end{equation}
\begin{equation}
    \label{Gaussian state}
    \ket{\psi}_{G} = \frac{\sum_{k=0}^N e^{-\rho(k-N/2)^2+isk\pi/2}\ket{N-k,k}}{\sqrt{\sum_{k'=0}^N (e^{-\rho(k'-N/2)^2})^2}} \,.
\end{equation}

For a given number of photons, Gaussian states are found to be optimal in the regime of high prior uncertainty and N00N states are optimal in the low prior uncertainty regime. In between these two regimes is a third intermediate regime whose optimal input states have complicated structures lacking a simple analytical expression~\cite{LK_Preprint}.

Gaussian states can be seen as a generalization of a type of coherent state called the Generalized Perelomov Coherent State (GPCS)~\cite{GPCS}. For coherent states, the variance of the photon number distribution is fixed, whereas here it is allowed to vary (the photon number follows a Super-Poissonian distribution with width depending on $\Delta$). Additionally, Gaussian and N00N inputs satisfy the $F>N$, condition for useful entanglement~\cite{useful_entanglement-Pezze}, and provide sub-shot-noise precision. F here is the Fisher information.

 Quasi-Gaussian states were introduced to model the optimal entangled photon states in the regime of large initial phase uncertainty $\Delta$~\cite{LK_Preprint}. Here we improve upon this earlier work  by showing that Gaussian states work almost equally well, while requiring fewer optimization variables.
 
A quasi-Gaussian state in Fock notation takes the following form \cite{LK_Preprint}:
\begin{equation}
    \label{quasi-Gaussian state}
    \ket{\psi}_{qG} = \frac{\sum_{k=0}^N e^{-\rho(k-N/2)^2-\rho'(k-N/2)^4+isk\pi/2}\ket{N-k,k}}{\sqrt{\sum_{k'=0}^N (e^{-\rho(k'-N/2)^2-\rho'(k'-N/2)^4})^2}} \,,
\end{equation}
where $s=\pm 1$ as before and $\rho, \rho'$ are functions of $N$ and $\Delta$. A pure Gaussian lacks the quartic correction term and corresponds to $\rho'=0$ in \eqref{quasi-Gaussian state}. Finding the optimal Gaussian for given $N,\,\Delta$ amounts to minimizing \eqref{eq:BMSE2} with respect to $\rho$ only (as opposed to optimizing over $\rho$ and $\rho'$ in the case of a quasi-Gaussian).

\begin{figure}[h]
    \includegraphics[width=\columnwidth]{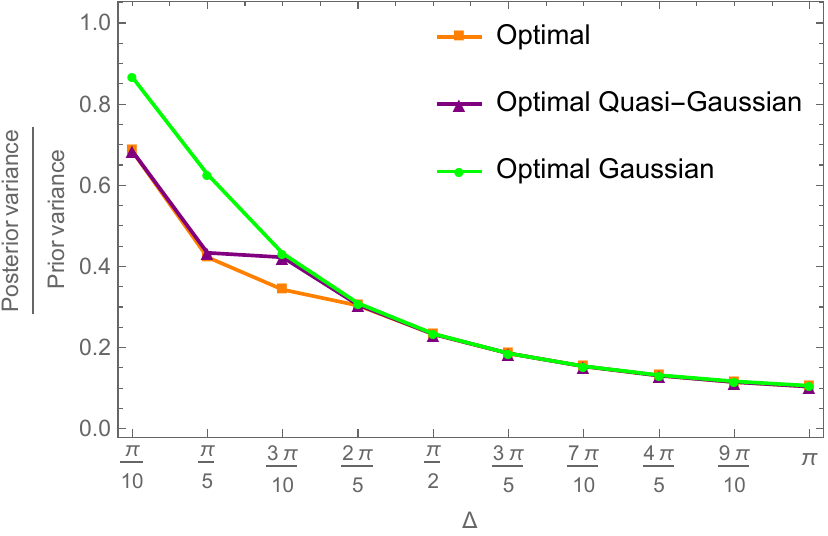}
    \caption{Comparing the ratio $\frac{\text{posterior variance}}{\text{prior variance}}$ for optimal 1-shot states (N=7). The result for the \textcolor{green}{optimal Gaussian (green)} is compared with the \textcolor{violet}{optimal quasi-Gaussian (violet)} input state as a function of the prior uncertainty $\Delta$. The \textcolor{orange}{globally optimal variance (orange)} is also plotted for comparison.}
    \label{fig:QG_G comparison}
\end{figure}

Even though Ref.~\cite{LK_Preprint} shows some examplss where the quasi-Gaussian state better represents the optimal state (e.g., for $N=11$, $\Delta=0.8\pi$ and $N=12$, $\Delta=0.9\pi$), we see in  Fig.~\ref{fig:QG_G comparison} that the posterior variance is very minimally affected. In this Figure, we plot $\frac{(\delta\phi_{posterior})^2}{(\delta\phi_{prior})^2}$ as a function of $\Delta$. Here $(\delta\phi_{posterior})^2$ is the BMSE defined in \eqref{eq:BMSE2} minimized over the possible input states. For a flat prior phase uncertainty, the variance $(\delta\phi_{prior})^2 = \Delta^2/12$. \\

By optimizing over special input states of the form  \eqref{Gaussian state} and \eqref{quasi-Gaussian state} instead of over the full space of possible input states, we reduce the number of optimization variables for a single shot from $2 (N+1)$ to just one variable ($\rho$) and two variables ($\rho,\rho'$) in the Gaussian and quasi-Gaussian cases respectively.
As expected, we see that Gaussian states are optimal for large uncertainties. We also see that the ratio of the posterior to prior variance is quite robust to small changes in the input states (e.g., quasi-Gaussian vs. Gaussian). We will use this robustness to our advantage, and in what follows use the simpler pure Gaussian states in the high prior uncertainty regime, so that only a single parameter $\rho$ needs to be optimized.

In the low prior uncertainty regime, where N00N states are optimal, Gaussian states do not work well. We note that in Fig.~\ref{fig:QG_G comparison}, the quasi-Gaussian states do appear to work well in the small $\Delta$ regime. This, however, is due to a mathematical curiosity: for $N$ odd and $\rho$ negative and large, the Gaussian state \eqref{Gaussian state} mimics the N00N state, and the same holds for the quasi-Gaussian state \eqref{quasi-Gaussian state} when either $\rho$ or $\rho'$ is negative and sufficiently large. Now in the case of the Gaussian, the physically motivated optimization constraint $\rho>0$ eliminates this potential mimicry, while in the case of the quasi-Gaussian no such obvious physically motivated constraint exists for the optimization. Thus the quasi-Gaussian state does appear to work well for small $\Delta$ in Fig.~\ref{fig:QG_G comparison}, but only because it is mimicking the N00N state.

Now for a single shot, Heisenberg’s uncertainty principle gives a fundamental limit to the accuracy with which $\delta\phi_{posterior}$ can be predicted. With respect to photon number, the lower bound scales asymptotically for large photon number $N$ as~\cite{N00N_states_optimal,HL}
\begin{equation}
    \label{Heisenberg uncertainty}
    \delta\phi_{min} \propto \frac{1}{N} \,.
\end{equation}
Indeed the N00N states obey this scaling, consistent with~\cite{HL, Dowling_N00N, Rubio2019, Rubio2020, Sidhu_Kok_review}; however, as seen in  Fig.~\ref{Boundaries}, the N00N states are effective only when the initial phase uncertainty $\Delta$ is already $O(1/N)$ or smaller~\cite{Rubio2019}. Physically, this is evident from the symmetry property $p(m_1|\phi+2\pi/N)=p(m_1|\phi)$ for a N00N input state, i.e. N00N states are only sensitive to the phase modulo $2\pi/N$. Indeed, a single-shot N00N measurement only decreases the phase uncertainty by a small amount, as discussed quantitatively in Sec.~\ref{N00N Uncertainty Scaling} below. We can further improve our knowledge about the phase by making multiple independent measurements.

\section{Extension to Multiple Shots -- Non-Adaptive Measurement Strategy}\label{Non-Adaptive formalism}
\begin{figure}[h]
\includegraphics[width=\columnwidth]{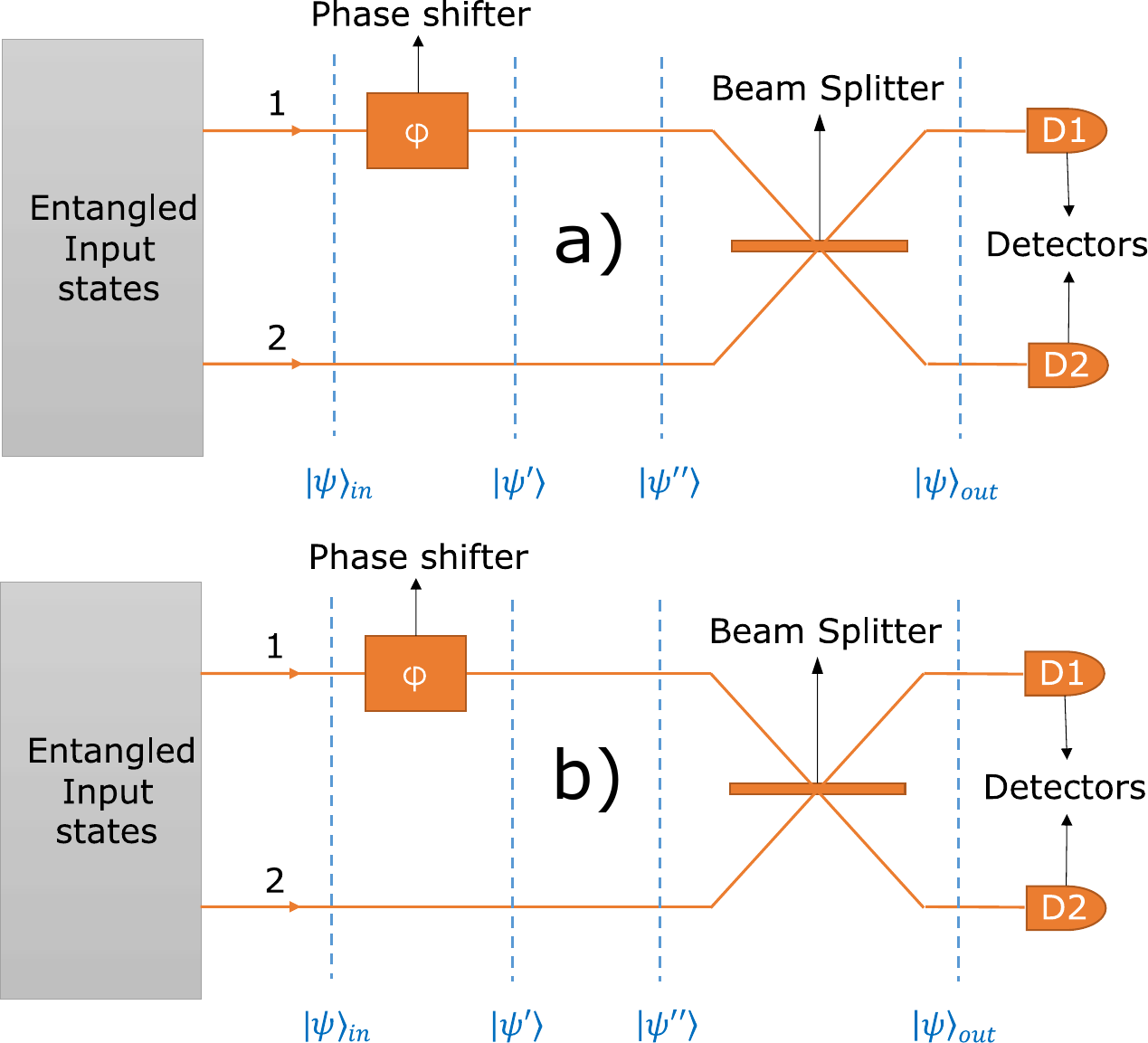}
\caption{Two-shot experiment schematic. a) Perform $1^{st}$ shot, gather $p(m_1|\phi)$ for each of the possible outcomes $m_1$. b) Perform $2^{nd}$ shot, gather $p(m_2|\phi)$ for each of the possible outcomes $m_2$.}
\label{fig:2ShotsSchematic}
\end{figure}

For a sequence of $\nu=2$ independent measurements  $m_1$, $m_2$, Fig.~\ref{fig:2ShotsSchematic} can be useful in visualizing the experiment, and can easily be extended to an arbitrary number of shots. We note that in contrast with the analysis in~\cite{Rubio2019}, we do not assume the input states for shots 1 and 2 to be identical.\\

In the non-adaptive approach discussed in this Section, we make a measurement after every shot, but do not use the outcome of the experiment to determine the best input for the next shot, Instead, we simply store the conditional PDFs $p(m_i|\phi)$ after each shot. 

Then, after we finish all $\nu$ shots, we gather all these conditional PDFs and construct our multi-shot BMSE. This BMSE is a function of all the input states for all $\nu$ shots. We can then minimize the BMSE globally over all $\nu$ input states (Sec.~\ref{Global Optimization}) or locally, i.e., shot by shot (Sec.~\ref{SBS}). Note that the local optimization method does take into account our knowledge that we will be performing multiple measurements. It is also more scalable for larger $N$ and $\nu$. 

Since the outcomes for different shots are statistically independent, i.e., the probability of a given outcome for one shot depends only on the true value of the phase and the input state {\it for that shot}, the single-shot probability \eqref{Pm}, the phase estimator \eqref{eq:phitilde} and the BMSE \eqref{eq:BMSE2} generalize to:

\begin{equation}
\label{eq:Pm multishot}
    p(m_1,m_2,...m_{\nu}) = \int_{-\Delta/2}^{\Delta/2} p(\phi) \prod_{i=1}^{\nu} p(m_i|\phi)  \, d\phi\
\end{equation}

\begin{equation}
    \tilde{\phi}_{m_1 m_2...m_{\nu}} = \frac{\int_{-\Delta/2}^{\Delta/2}\phi\,p(\phi)\prod_{i=1}^{\nu} p(m_i|\phi) \, d\phi\ }{p(m_1,m_2,...m_{\nu})}
    \label{eq: phitilde multishot}
\end{equation}

\begin{multline}
    \label{eq:BMSE multishot}
    BMSE(\tilde{\phi}_{m_1 m_2...m_{\nu}}) =\\
    \sum_{m_1 m_2...m_{\nu}} \int_{-\Delta/2}^{\Delta/2} (\phi - \tilde{\phi}_{m_1 m_2...m_{\nu}})^2 p(\phi) \prod_{i=1}^{\nu} p(m_i|\phi) \, d\phi \   \,.
\end{multline}

The BMSE here is summed over all possible outcome sequences $m_1 m_2...m_{\nu}$. 

\subsection{Non-Adaptive Global Optimization}
 \label{Global Optimization}
 
 \begin{figure}[h]
    \centering
    \captionsetup{singlelinecheck=off}
    	\includegraphics[width=\columnwidth]{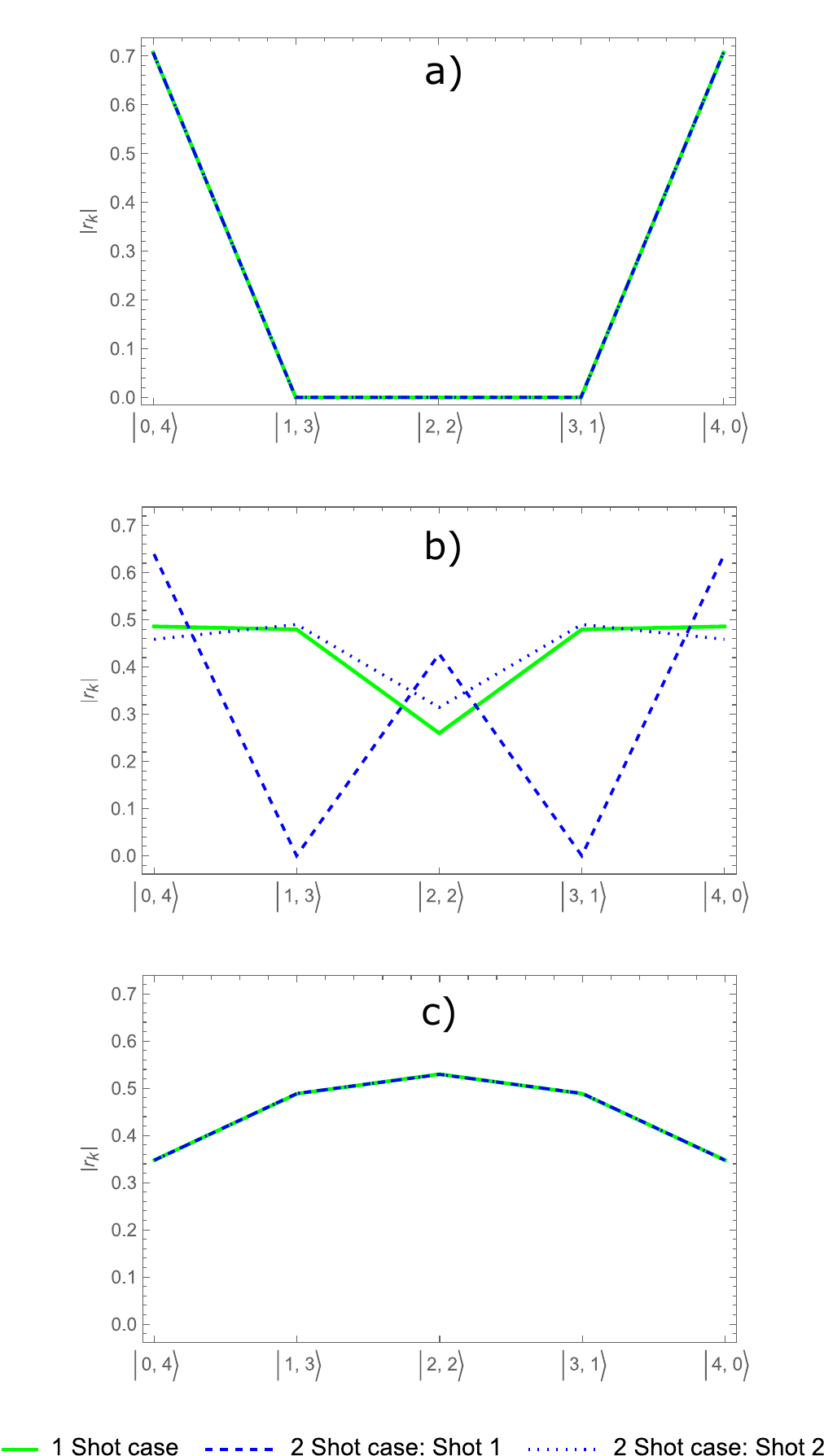}
		\caption{Comparing the two 2-shot optimal input states with the single 1-shot optimal input state, for $N=4$ photons and three different values of the prior uncertainty corresponding to the three regimes:
	a) $\Delta=\frac{\pi}{10}$ (N00N regime). b) $\Delta=\frac{\pi}{2}$ (Intermediate regime). c) $\Delta=\pi$  (Gaussian regime).}
    \label{fig:2Shots}
\end{figure}

In the three panels of Fig.~\ref{fig:2Shots}, we compare the 2-shot optimal input state amplitudes with the single-shot optimal input state for $N=4$ photons and for three different values of the prior uncertainty, corresponding to the three regimes: N00N, intermediate, and Gaussian.

We see that the optimal two-shot input states match the optimal one-shot input states for the N00N and Gaussian regimes almost perfectly. For the case where the initial uncertainty $\Delta$ is in the intermediate regime, Fig.~\ref{fig:2Shots}(b), the one-shot optimization approximates but does not exactly match one of the two states obtained in the global two-shot optimization. As we mentioned earlier, the intermediate state is quite difficult to analyze. Since the intermediate regime is a narrow one and can usually be traversed in a single shot~\cite{LK_Preprint}, the analytical intractability of this regime is not a major concern.

The same optimization protocol can be extended to $\nu$ shots and is only limited by a computer's processing capacity. Our results are consistent with the conjecture that in a global $\nu$-shot optimization for general $\nu$, at least one of the $\nu$ optimal input states will also match the one-shot optimal input state.

\subsection{Analytical Formulae for N00N and Gaussian Inputs}\label{Analytical formulae}
A difficulty we face while determining the optimal input states is that the number of optimization variables $r_k$ and $\theta_k$ increases as $2\nu(N+1)$, or $2\nu N$ using normalization and the irrelevance of the overall phase for each input state. Using the fact that in the lossless case the optimal input states are symmetric w.r.t. mode interchange~\cite{LK_Preprint}
\begin{equation}
    r_k=r_{N-k} \geq 0, \;\;\; \theta_k=-\theta_{N-k} \,,
\end{equation}
allows us cut the number of optimization variables in half. Even then, brute force optimization by numerically minimizing the variance over $\nu N$ variables quickly becomes unfeasible for larger systems (more photons or more shots). For this reason, we look for analytical formulae that best approximate the optimal input states.

\begin{figure}[h]
    \centering
    	\includegraphics[width=\columnwidth]{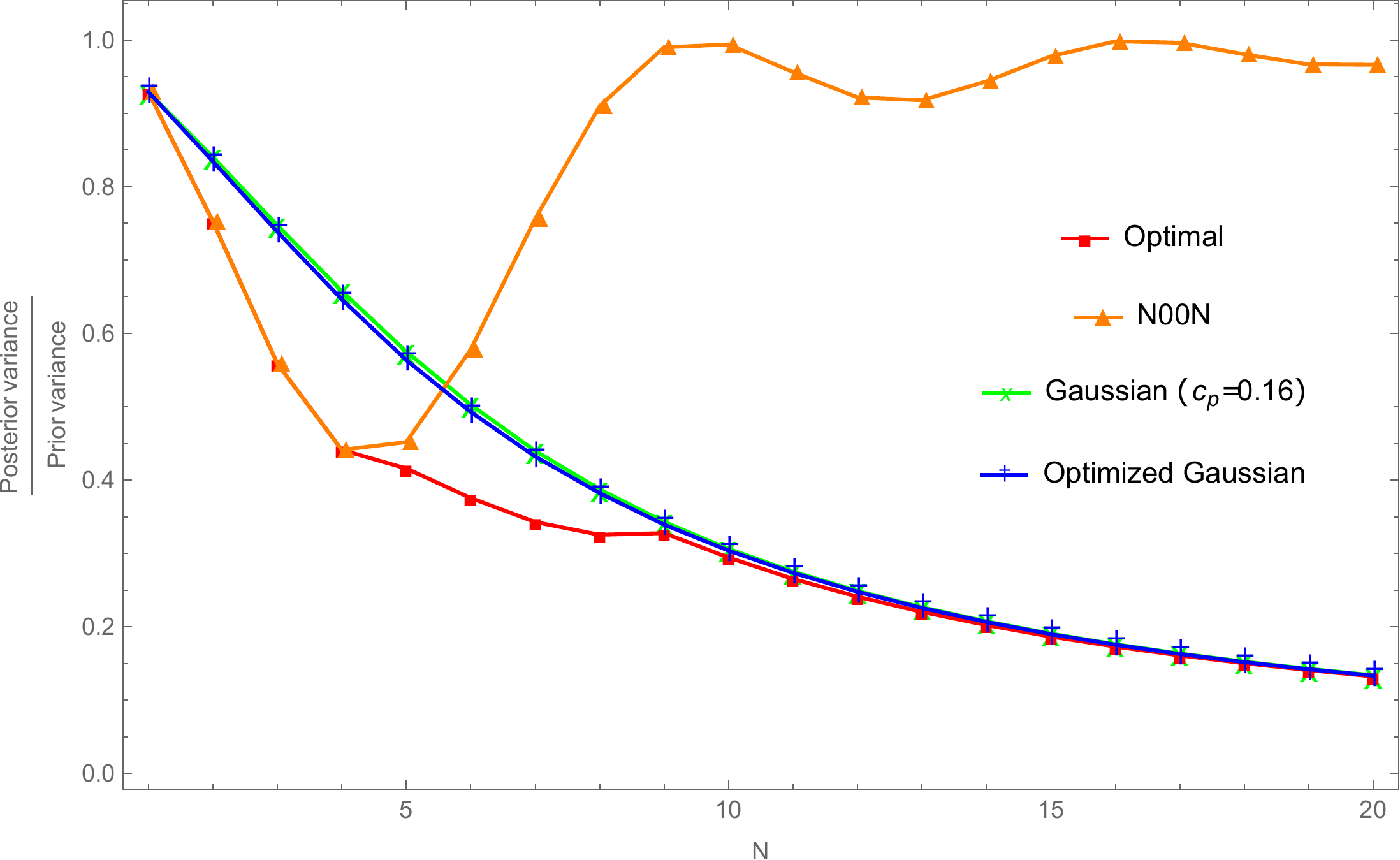}
		\caption{$\frac{\text{posterior variance}}{\text{prior variance}}$ as a function of photon number $N$ is obtained for $\Delta=3\pi/10$ using several different input states.}
    \label{fig:best-fit states}
\end{figure}

Since the N00N input state as given analytically by \eqref{N00N} is already parameter-free, we need only to determine the optimal input in the Gaussian regime, i.e., an optimal value for $\rho$ in \eqref{Gaussian state}. We make use of the scaling $\rho = c_\rho\frac{\Delta}{N}$~\cite{LK_Preprint} and find that the model Gaussian with $c_\rho \approx 0.16$ is a good approximation to the optimal Gaussian.
In Fig.~\ref{fig:best-fit states} we plot the ratio of prior variance to posterior variance,
$(\delta\phi_{posterior})^2/(\delta\phi_{prior})^2$ for $\Delta=3\pi/10$ as a function of $N$ and see that the result using this simple analytical formula \textcolor{green}{(green, x)} matches almost perfectly with the result of numerical optimization \textcolor{blue}{(blue, dagger)}. This is in part due to the robustness of the variance to small changes in the input states, which we use to our advantage. 

 The optimal Gaussian is then expressed as follows:
 \begin{equation}
    \ket{\psi}_{G} = \frac{\sum_{k=0}^N e^{-0.16\frac{\Delta}{N}(k-N/2)^2+isk\pi/2}\ket{N-k,k}}{\sqrt{\sum_{k'=0}^N (e^{-0.16\frac{\Delta}{N}(k'-N/2)^2})^2}} \,.
\end{equation}

\begin{figure}[h]
    \centering
    \includegraphics[width=\columnwidth]{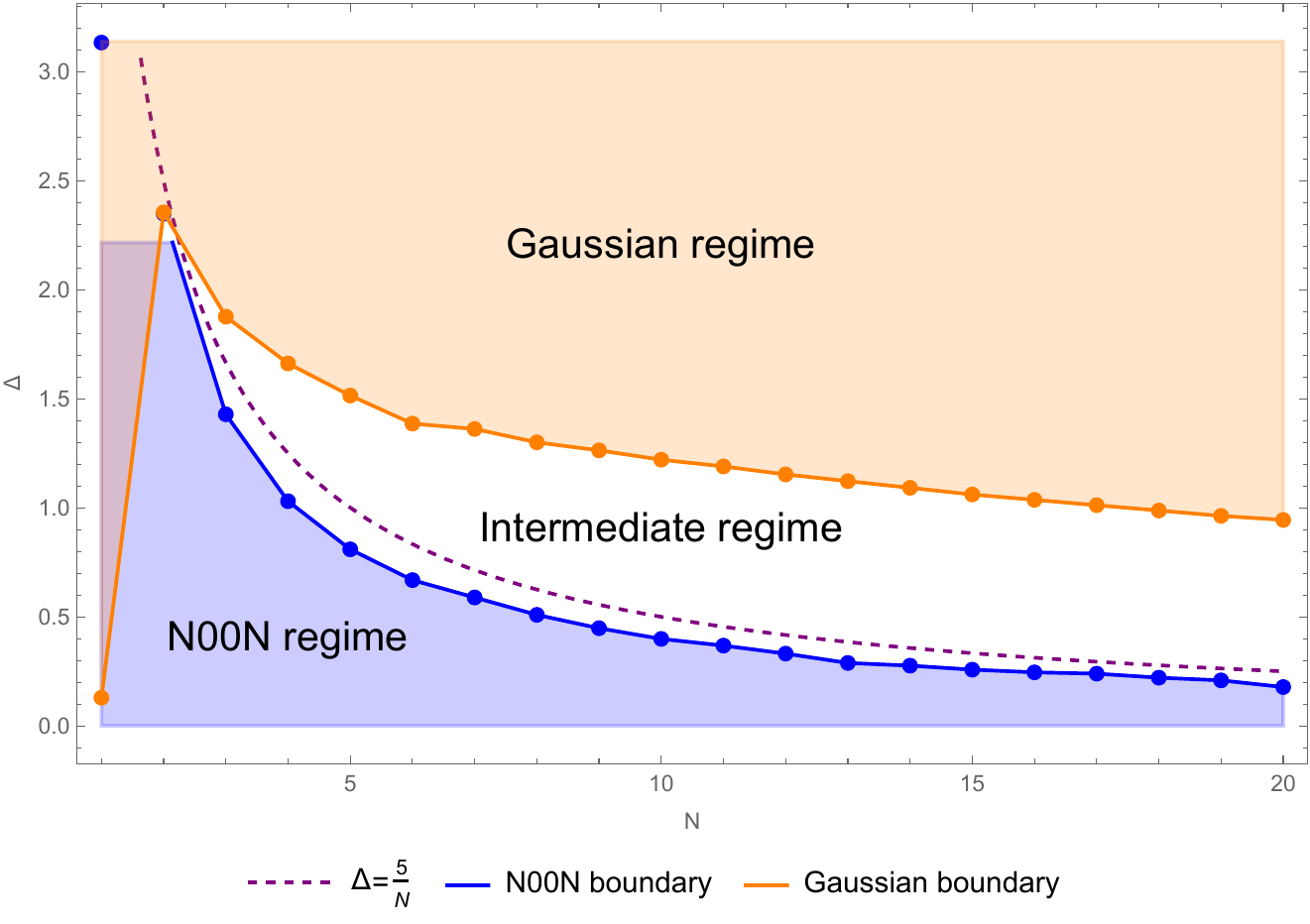}
    \caption{The optimal input state regimes are shown for different prior uncertainties $\Delta$ and number of photons $N$. Gaussian or N00N input states are optimal when $(N,\Delta)$ is a point in the orange or blue region respectively.}
    \label{Boundaries}
\end{figure}

 Now that we determined analytical expressions for N00N and Gaussian inputs, we would like to bypass the variance optimization step entirely and use these analytical expressions as inputs in their respective regimes. The problem now becomes one of determining boundaries between these regimes, and when it is appropriate to use which state. This becomes a classification problem. For a fixed $N$, the boundary between the regimes may be determined numerically using the bisection method and the resulting phase diagram is shown in Fig.~\ref{Boundaries}.
 
Note that the separation into regimes breaks down for the single photon ($N=1$) case since there the optimal state is neither/both a N00N and a Gaussian. Ref.~\cite{LK_Preprint} suggests that the Gaussian regime can only hold for $N\Delta \geq 1$, which is in conformance with our results here.

For the purpose of developing a simple rule to determine the aptness of an input state, we decide to include the intermediate regime inside the Gaussian regime. Even though Gaussian input states are not optimal in the intermediate regime, they can still provide a significant reduction in uncertainty in that regime, as can be seen in Fig.~\ref{fig:best-fit states}. Therefore, in the following we will use the simple $N\Delta = 5$ boundary to differentiate the N00N regime ($N\Delta<5$) from the Gaussian/intermediate regime ($N\Delta>5$). We will see more use of this demarcation in Sec.~\ref{Scaling}.

\subsection{Non-Adaptive Local Optimization} \label{SBS}
In the global optimization approach of Sec.~\ref{Global Optimization}, $\nu$ measurements are performed, but all $\nu$ input states are predetermined in advance through a global optimization process. Here, we simplify the analysis by performing the optimization locally (one shot at a time). The actual outcomes of the previous shots are still not used. Instead, the posterior variance averaged over all possible outcomes $m$ of the previous shot is used to set the prior variance of the next shot. Explicitly, we do the following:
\begin{enumerate}
    \item Assume a flat prior uncertainty for the $n^{th}$ shot
    \item Find $(\delta\phi_{posterior})^2$, averaged over all possible measurement outcomes \eqref{eq:BMSE2}
    \item Find the optimal input state that minimizes $(\delta\phi_{posterior})^2$
    \item The $n^{th}$ shot posterior uncertainty becomes the prior uncertainty for the $(n+1)^{st}$ shot
    \item Repeat above steps for each shot, $n=1\ldots \nu$
\end{enumerate}

\begin{figure}[h]
    \centering
    \includegraphics[width=0.5\textwidth]{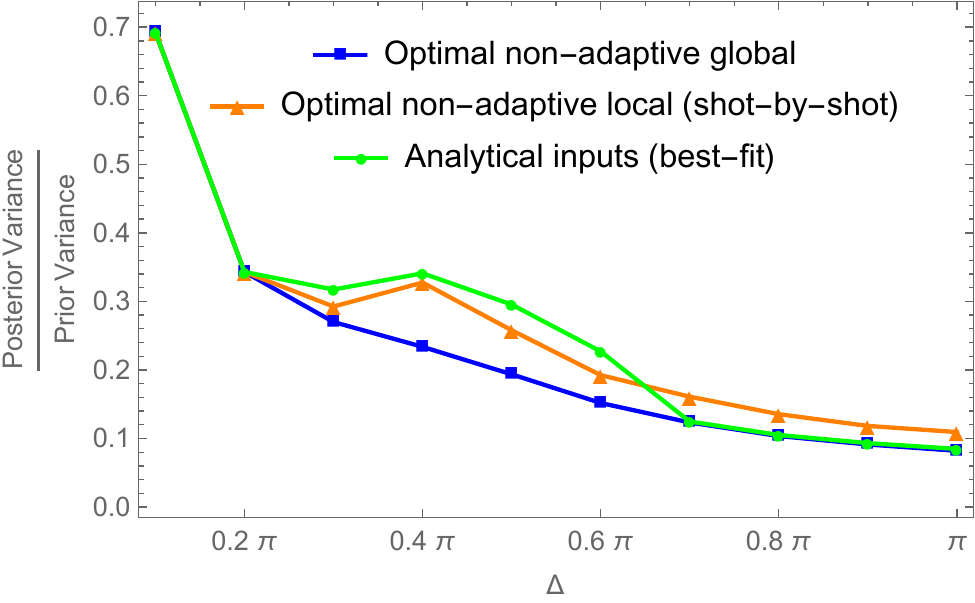}
    \caption{$\frac{\text{posterior variance}}{\text{prior variance}}$ for various values of the prior uncertainty $\Delta$, using two shots with five photons each ($\nu=2$, $N=5$).}
    \label{Feedback comparison}
\end{figure}

In Fig.~\ref{Feedback comparison} we analyze for the case of $\nu=2$ shots the ratio of posterior to prior variance as a function of the prior uncertainty $\Delta$ (analogous to Fig.~\ref{fig:QG_G comparison} in the single-shot scenario). The \textcolor{blue}{(blue, square)} line in Fig.~\ref{Feedback comparison} corresponds to the non-adaptive, two-shot,
globally optimized variance ratio from Sec.~\ref{Global Optimization}. The \textcolor{orange}{(orange, triangle)} line corresponds to the optimized variance ratio outlined in the steps above, where we optimize the variance one shot at a time. The  \textcolor{green}{(green, circle)} line corresponds to the optimized variance ratio, where we use the best-fit states according to the formulae shown in Sec.~\ref{Analytical formulae}.

In the global non-adaptive optimization case \textcolor{blue}{(blue, square)}, we have to optimize the variance with respect to $2\nu (N+1)$ variables at one time. In comparison, in the shot-by-shot non-adaptive case \textcolor{orange}{(orange, triangle)},
we have only $2(N+1)$ variables to deal with per shot.
There is an important approximation that we make in the shot-by-shot case, which may not always be valid. The global optimization does not make any assumptions about the shape of the probability distribution prior to the second shot. However, in the local optimization (shot-by-shot) case, we assume that this probability distribution prior to the second shot is again flat, albeit narrower than the probability distribution prior to the first shot, This approximation causes the discrepancy between the orange and blue lines in Fig.~\ref{Feedback comparison}, noticeable primarily in the intermediate regime.

We also make use of our best-fit models from Sec.~\ref{Analytical formulae}
to obtain the variance shown in \textcolor{green}{(green, circle)} in Fig.~\ref{Feedback comparison}. This method is by far the cheapest computationally, and also provides us with a very close approximation to the more expensive optimization methods. The difference between the three methods in the interval $\frac{2\pi}{10}\leq\Delta\leq \frac{7\pi}{10}$ is caused by the unpredictable intermediate states. Unsurprisingly, the simple analytical formulae perform the worst in this regime. From Fig.~\ref{Feedback comparison} we see that even though we make an unphysical assumption about the prior probability distribution for the second shot (assuming this distribution is flat), the analytical model from Sec.~\ref{Analytical formulae} performs very well in the N00N and Gaussian regimes.

\subsection{Non-Adaptive Local Scaling} \label{Scaling}
Using the results in Sections~\ref{Global Optimization}, \ref{Analytical formulae}, and \ref{SBS}, we try to predict how much information can expect to gain per shot, on average. We determine how the uncertainty of phase estimation scales with the number of shots. Specifically, our goal in this Section is to determine how many shots we would need on average to arrive at a required uncertainty $\Delta_{req}$ given a starting uncertainty $\Delta_{start}$. To make the analysis tractable for large number of shots $\nu$, we take the optimal states to be determined shot-by-shot as in
Sec.~\ref{SBS} with the optimal state at each shot obtained using the analytical formulae obtained in Sec.~\ref{Analytical formulae}.

\textbf{Nomenclature:}
\begin{itemize}
    \item $\Delta_{in},\Delta_{out}:$ Input uncertainty and output uncertainty after a single shot regardless of regime (Single shot only, may cross regime boundaries)
    \item $\Delta_{i},\Delta_{f}:$ Initial uncertainty and final uncertainty after multiple shots constrained to a single regime (Regime specific and does not cross regime boundaries)
    \item $\Delta_{start},\Delta_{req}:$ Starting and required uncertainties regardless of regime (May require multiple shots, may cross regime boundaries)
\end{itemize}

\subsubsection{Gaussian Uncertainty Scaling}
In the Gaussian regime, the posterior uncertainty after one shot follows the scaling
\begin{equation}
    \label{Gaussian scaling formula}
    \Delta_{out} \approx c_G \frac{\sqrt{\Delta_{in}}}{\sqrt{N}} \,,
\end{equation}
where numerically we find the parameter $c_G$ to take the value $c_G \approx 1.27$.
Then the uncertainty for the $(n+1)^{st}$ shot is given as a function of the $n^{th}$ shot uncertainty:
\begin{equation}
    \Delta_{n+1} = c_G \sqrt{\frac{\Delta_n}{N}} \,.
    \label{iterategaussian}
\end{equation}
Given the initial uncertainty $\Delta_0 = \Delta_i$, iterating \eqref{iterategaussian} $\nu$ times yields the final uncertainty  $\Delta_{\nu} = \Delta_f$. Again assuming that $\Delta_i,\Delta_f$ both reside in the Gaussian regime, we have:
 \begin{equation}
     \Delta_f = \Delta_i^{2^{-\nu}}\exp\bigg((1-2^{-\nu})(2\ln c_G - \ln N)\bigg) \,.
 \end{equation}
 
 Solving for $\nu_G$,  the number of shots required to get from $\Delta_i$ to $\Delta_f$ (provided both reside in the Gaussian regime), we obtain
 \begin{equation}
     \label{NumShotsreq Gaussian}
     \nu_G = \Bigg\lceil\frac{\ln\bigg(\frac{\ln(N\Delta_i/c_G^2)}{\ln(N\Delta_f/c_G^2)}\bigg)}{\ln 2}\Bigg\rceil \,,
 \end{equation}
 where $\lceil \ldots \rceil$ represents the ceiling function.

\subsubsection{N00N Uncertainty Scaling}\label{N00N Uncertainty Scaling}

Far in the N00N regime ($\Delta_{in} \ll 1/N$), the reduction in uncertainty in one shot is relatively small. Indeed, Fig.~\ref{fig:Scaling delta,N dependence}~a) shows that  $\Delta_{out}-\Delta_{in} \propto -\Delta_{in}^3$ for small $\Delta_{in}$. We let $\Delta_{out}-\Delta_{in} = - c_N(N) \Delta_{in}^3$, where $c_N(N)$ is an $N-$dependent coefficient, and examining the behavior as a function of $N$ for fixed $\Delta_{in}$ in Fig.~\ref{fig:Scaling delta,N dependence}~b) we obtain the scaling
$c_N(N) \propto -N^2$.

\begin{figure}[h]
    \centering
    \includegraphics[width=0.3\textwidth]{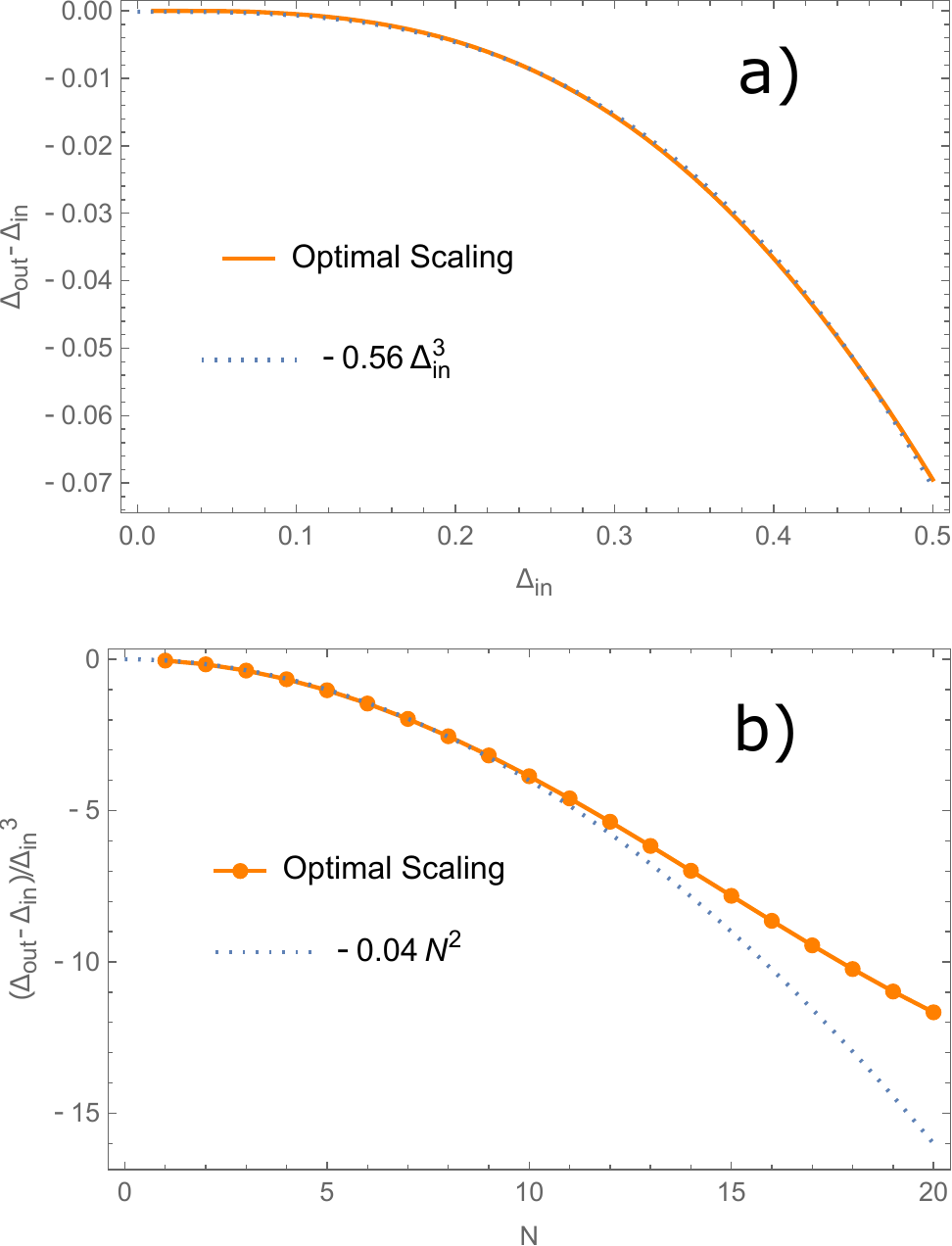}
    \caption{The uncertainty reduction $\Delta_{out}-\Delta_{in}$ as a function of $\Delta_{in}$ and $N$ far in the N00N regime, for one shot. a) $\Delta_{out}-\Delta_{in} \propto \Delta_{in}^3$ for $N=4$;
    b) $\Delta_{out}-\Delta_{in} \propto N^2$ for $\Delta_{in} = \pi/20$.}
    \label{fig:Scaling delta,N dependence}
\end{figure}

Therefore, in the N00N regime, we find the one-shot uncertainty reduction to be
\begin{equation}
    \label{N00N scaling formula}
    \Delta_{out} - \Delta_{in} \approx - c_N N^2 {\Delta_{in}^3} \,, 
\end{equation}
where the numerical coefficient takes the value $c_N  \approx 0.04$.

In the deep N00N regime $N \Delta_{in} \ll 1$, $|\Delta_{out} - \Delta_{in}| \ll \Delta_{in}$, so we may rewrite 
the difference equation \eqref{N00N  scaling formula} for one shot in differential form:
\begin{equation}
    d\Delta = -c_N\;N^2\;\Delta^3\;dn \,.
\end{equation}
Integrating from $n=0$ to $\nu$, we have
\begin{equation}
    \frac{1}{\Delta_f^2} - \frac{1}{\Delta_i^2} = 2c_N\;N^2\;\nu + 0 \,, \label{N00N scaling continuous}
\end{equation}
where $\nu$ is the number of shots and the integration constant is fixed at 0 by the initial condition $\Delta_f=\Delta_i$ for $\nu=0$. Therefore the number of shots required to reduce the phase uncertainty from $\Delta_i$ to $\Delta_f$, provided that both reside in the N00N regime, is given by the ceiling function
\begin{equation}
\label{NumShotsreq N00N}
     \nu_N= \Bigg\lceil\left(\frac{1}{\Delta_f^2} - \frac{1}{\Delta_i^2}\right) \frac{1}{2c_N\;N^2}\Bigg\rceil \,.
\end{equation}

While the focus of the present work is to investigate the information gain for a finite number of shots, we remark that as $\nu$ becomes very large, \eqref{N00N scaling continuous} reduces to the asymptotic form
 \begin{equation}
      \frac{1}{\Delta_f^2} = 2c_N\;N^2\;\nu \implies \Delta_f \propto \frac{1}{N}\frac{1}{\sqrt{\nu}} \,,
 \end{equation}
which is the scaling we expect in this limit due to the Heisenberg limit and the central limit theorem, in conformance with~\cite{Pezze}.

\begin{figure}[h]
    \centering
    \includegraphics[width=0.5\textwidth]{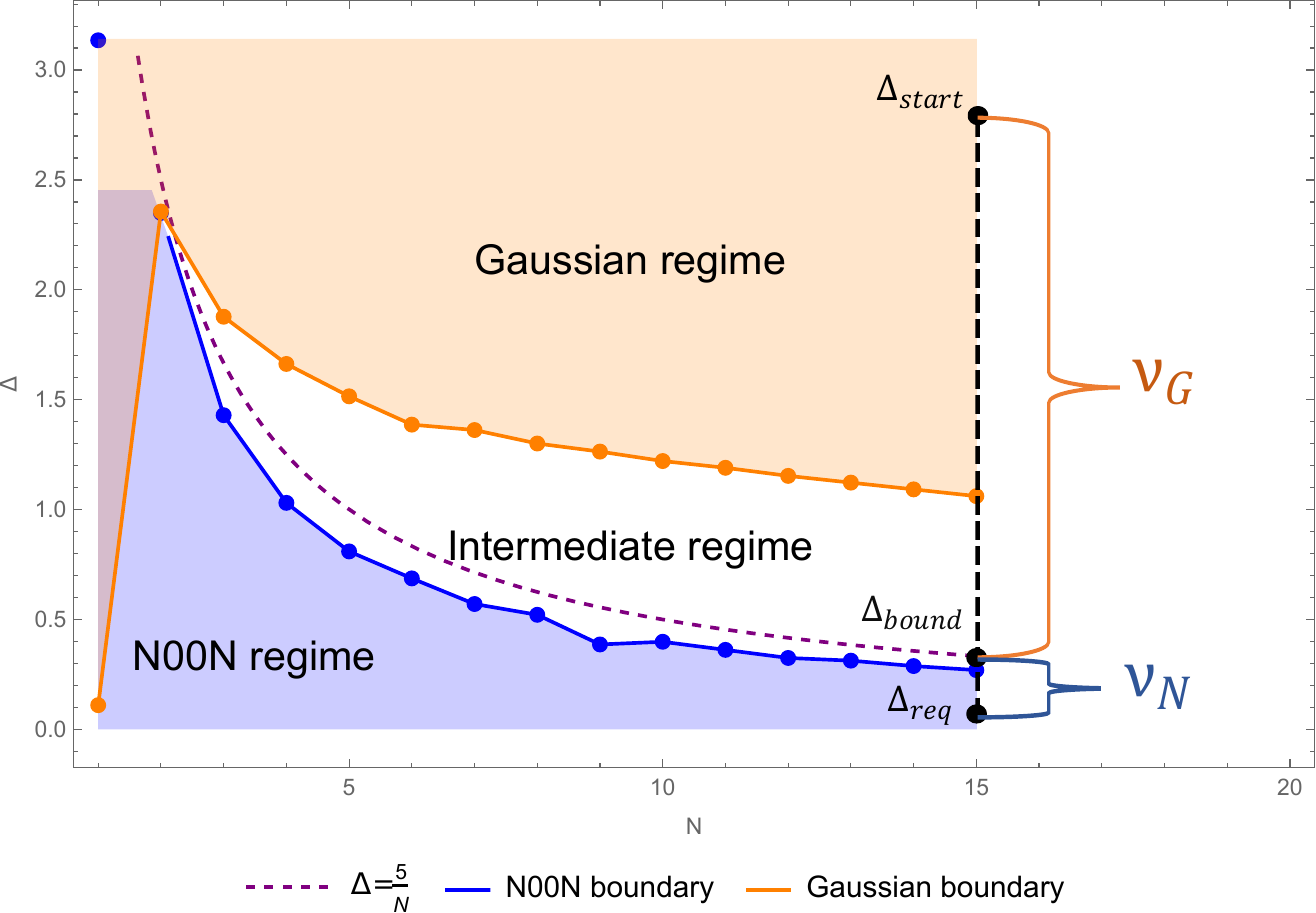}
    \caption{$\nu_G$ is the number of shots needed on average, using Gaussian inputs, to arrive at the boundary uncertainty $\Delta_{boundary}$ given a starting initial uncertainty $\Delta_{start}$. $\nu_N$ is the number of shots needed on average, using N00N inputs, to arrive at the required uncertainty $\Delta_{req}$ from $\Delta_{boundary}$.}
    \label{nsgnsn}
\end{figure}

\subsection{General Predictions}\label{General}
For a general initial uncertainty $\Delta_{start}$ and desired uncertainty $\Delta_{req}$, we first check the regimes in which $\Delta_{start}$ and $\Delta_{req}$ reside. When both are in the Gaussian or in the N00N regime, \eqref{NumShotsreq Gaussian} or \eqref{NumShotsreq N00N}, respectively may be applied directly. If $\Delta_{start}$ is in the Gaussian regime and $\Delta_{req}$ in the N00N regime, we must add (i) the steps required to get to the Gaussian-N00N boundary, given by \eqref{NumShotsreq Gaussian} with $\Delta_i=\Delta_{start}$ and $\Delta_f=\Delta_{boundary}=5/N$ and (ii) the steps required to traverse the N00N regime to the final desired precision, given by 
\eqref{NumShotsreq N00N} with $\Delta_i=\Delta_{boundary}$ and $\Delta_f=\Delta_{req}$. This process is illustrated schematically in Fig.~\ref{nsgnsn}.

We note that the asymptotic formulae \eqref{NumShotsreq Gaussian} and \eqref{NumShotsreq N00N} are both approximate, valid for $N \Delta \gg 1$ and $\Delta \ll N \Delta \ll 1$, respectively. In particular both approximations are expected to break down near the boundary, where $\Delta \sim 1/N$. To see how these predictions work quantitatively, in Table~\ref{General predictions} we compare the predictions with exact numerical results for several cases. In all cases considered, the predicted number of shots needed deviates at most by one from the actual number obtained numerically.

\begin{table}[h]
\caption{\label{General predictions}The number of shots needed to reduce the phase uncertainty from the initial value of $\Delta_{start}$ to a target uncertainty $\Delta_{req}$. The actual number of shots obtained numerically (``Opt'') is compared with the number predicted using the analytical formulae \eqref{NumShotsreq Gaussian} and \eqref{NumShotsreq N00N} (``Formula''). }
\begin{ruledtabular}
\begin{tabular}{cccccccc}
Regime&N&$\Delta_{start}$&$Regime_{st}$&$\Delta_{req}$&$Regime_{req}$&Opt&Formula\\
\hline
Gaussian&9&$\pi$&Gaussian&0.5&Gaussian&3&2\\
N00N&9&$\frac{\pi}{15}$&N00N&$\frac{\pi}{20}$&N00N&3&3\\
Mixed&9&$\pi$&Gaussian&$\frac{\pi}{20}$&N00N&8&4+3=7\footnote{4+3 corresponds to 4 shots in the Gaussian regime followed by 3 shots in the N00N regime.}\\
Mixed&13&$\pi$&Gaussian&$0.05$&N00N&30&4+27=31\footnote{4+27 corresponds to 4 shots in the Gaussian regime followed by 27 shots in the N00N regime.}\\
Mixed&9&$\pi$&Gaussian&$0.05$&N00N&62&4+59=63\footnote{4+59 corresponds to 4 shots in the Gaussian regime followed by 59 shots in the N00N regime.}\\
\end{tabular}
\end{ruledtabular}
\end{table}

\section{Adaptive Measurement Strategy} \label{Adaptive formalism}
In the non-adaptive global formalism (Sec.~\ref{Global Optimization}), all $\nu$ input states to be used for the $\nu$ measurements are determined up front in a global optimization. Then, in the non-adaptive local formalism (Sec.~\ref{SBS}), we assumed a flat prior probability distribution before every shot and used the total average posterior variance (averaged over all possible measurement outcomes $m$ of that shot) to estimate the prior uncertainty for the next shot.

In the adaptive formalism to be discussed in this Section, the idea is to make a sequence of adaptive measurements, where the input state for each measurement depends on the outcome of the previous shots.
Unlike in Sec.~\ref{SBS}, we do not average over the possible outcomes. Again, we can optimize globally (Sec.~\ref{Adaptive Global}) or locally (Sec.~\ref{Feedforward}).

\subsection{Adaptive Global Optimization}\label{Adaptive Global}
In the global optimization method, we do not make any assumptions about the shape of the posterior probability distributions. We make use of the Bayesian sequential updating formula~\cite{Kay}, which for two shots with outcomes $m_1, m_2$ takes the form
 \begin{equation*}
         p(\phi|m_1,m_2) = p(\phi|m_2,m_1) = \frac{p(\phi|m_1)p(m_2|\phi,m)}{p(m_2|m)} \,,
 \end{equation*}
where $p(\phi|m_1)= \frac{p(m_1|\phi) p(\phi)}{p(m_1)}$, and therefore the posterior probability distribution after two shots is given by
\begin{equation}
     p(\phi|m_1,m_2) = \frac{p(\phi) p(m_1|\phi)p(m_2|\phi,m_1)}{p(m_1)p(m_2|m_1)}  \,.\label{Bayes' seq upd for phi}
\end{equation}

Denoting by $p(m_2,m_1)$ the joint probability of outcome $m_2$ occurring in the $2^{nd}$
shot and outcome $m_1$ occurring in the $1^{st}$ shot, $p(m_2,m_1)$ is equal to the denominator of \eqref{Bayes' seq upd for phi}:
\begin{equation}
    p(m_2,m_1) = p(m_1) p(m_2|m_1)  \,. \label{Pm2m1}
\end{equation}

Rearranging \eqref{Bayes' seq upd for phi} and integrating over all $\phi$, we have:
\begin{equation}
    p(m_2,m_1) = \int_{-\frac{\Delta}{2}}^{\frac{\Delta}{2}} p(\phi) p(m_1|\phi)p(m_2|\phi,m_1) \, d\phi \,.
\end{equation}
We can now construct the mean phase estimator $\tilde{\phi}_{m_2,m_1}$ by taking the mean of the posterior PDF $p(\phi|m_1,m_2)$
 \begin{equation}
    \tilde{\phi}_{m_2,m_1} = \frac{\int_{-\frac{\Delta}{2}}^{\frac{\Delta}{2}} \phi\,p(\phi) p(m_1|\phi)p(m_2|\phi,m_1)\, d\phi\ }{p(m_2,m_1)} \,. \label{phitilde seq upd}
\end{equation}

Finally, we can construct the BMSE 
\begin{multline}
    BMSE_{adapt} = \sum_{m_2}p(m_2|m_1) \times \\
    \sum_{m_1} p(m_1) \frac{\int_{-\frac{\Delta}{2}}^{\frac{\Delta}{2}} (\phi - \tilde{\phi}_{m_2,m_1})^2 p(\phi) p(m_1|\phi)p(m_2|\phi,m_1)\, d\phi\ }{p(m_2,m_1)}  \,. \label{variance seq upd}
\end{multline}
We note that the adaptive expressions \eqref{Pm2m1} -- \eqref{variance seq upd} for the probability, the phase estimator, and the BMSE are analogous to
\eqref{eq:Pm multishot}, \eqref{eq: phitilde multishot}, and \eqref{eq:BMSE multishot} for the non-adaptive case, except that here $p(m_2|\phi)$ is replaced by $p(m_2|\phi,m_1)$ since the input state for the second shot depends on the outcome $m_1$ of the first shot measurement. The extension from $\nu=2$ to an arbitrary number of shots is obvious.

\begin{figure*}
\includegraphics[width=\textwidth]{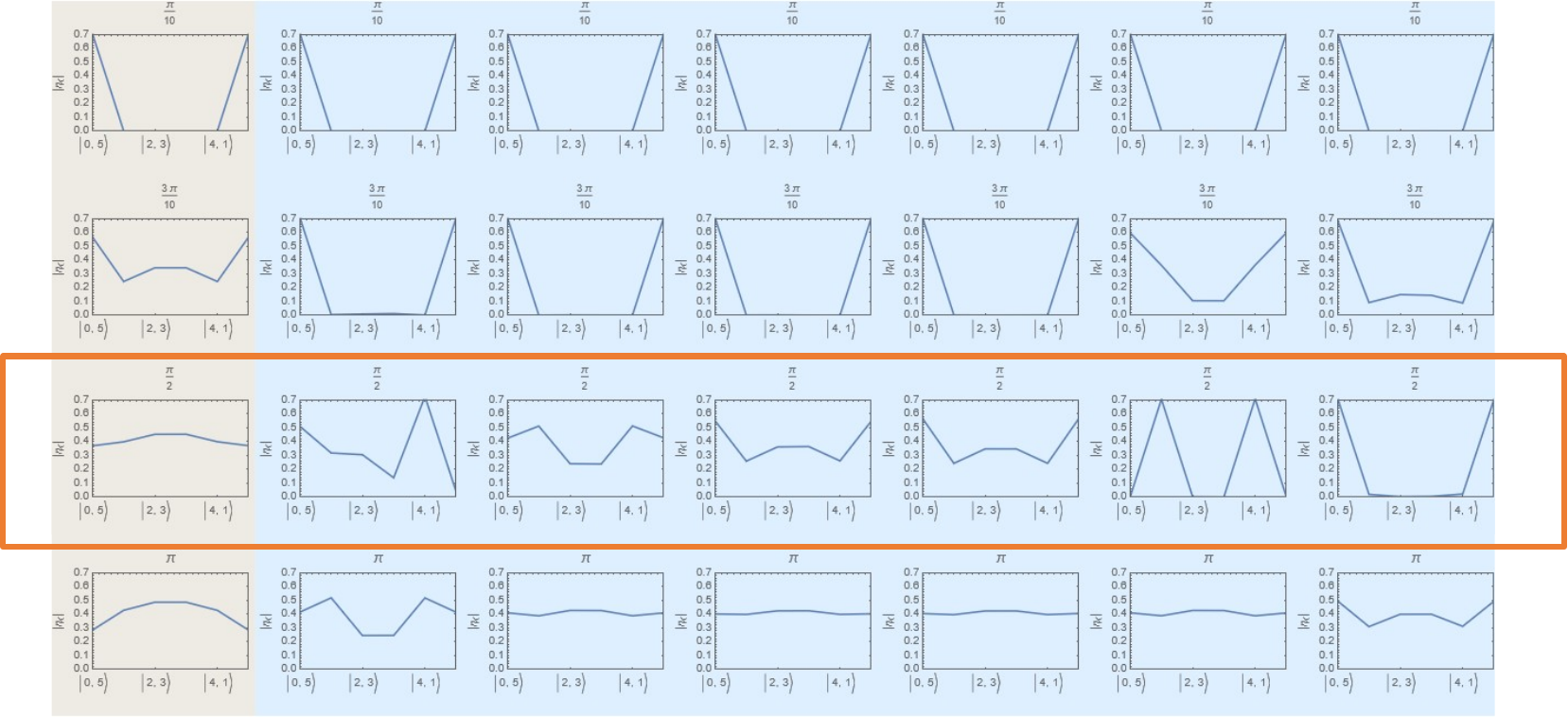}
\caption{\label{Adaptive global optimal state}Optimal inputs for $N=5$ photons, $\nu=2$ shots obtained via global adaptive optimization. The four rows, from bottom to top, show results
    for  prior uncertainties $\Delta=\pi,\frac{\pi}{2},\frac{3\pi}{10},\frac{\pi}{10}$, respectively. In each row, the leftmost state is the optimal input state to be used for the first-shot measurement, and the following six states are the six optimal second-shot states, one for each possible outcome of the first-shot measurement.}
\end{figure*}

\begin{figure*}
    \includegraphics[width=0.85\textwidth]{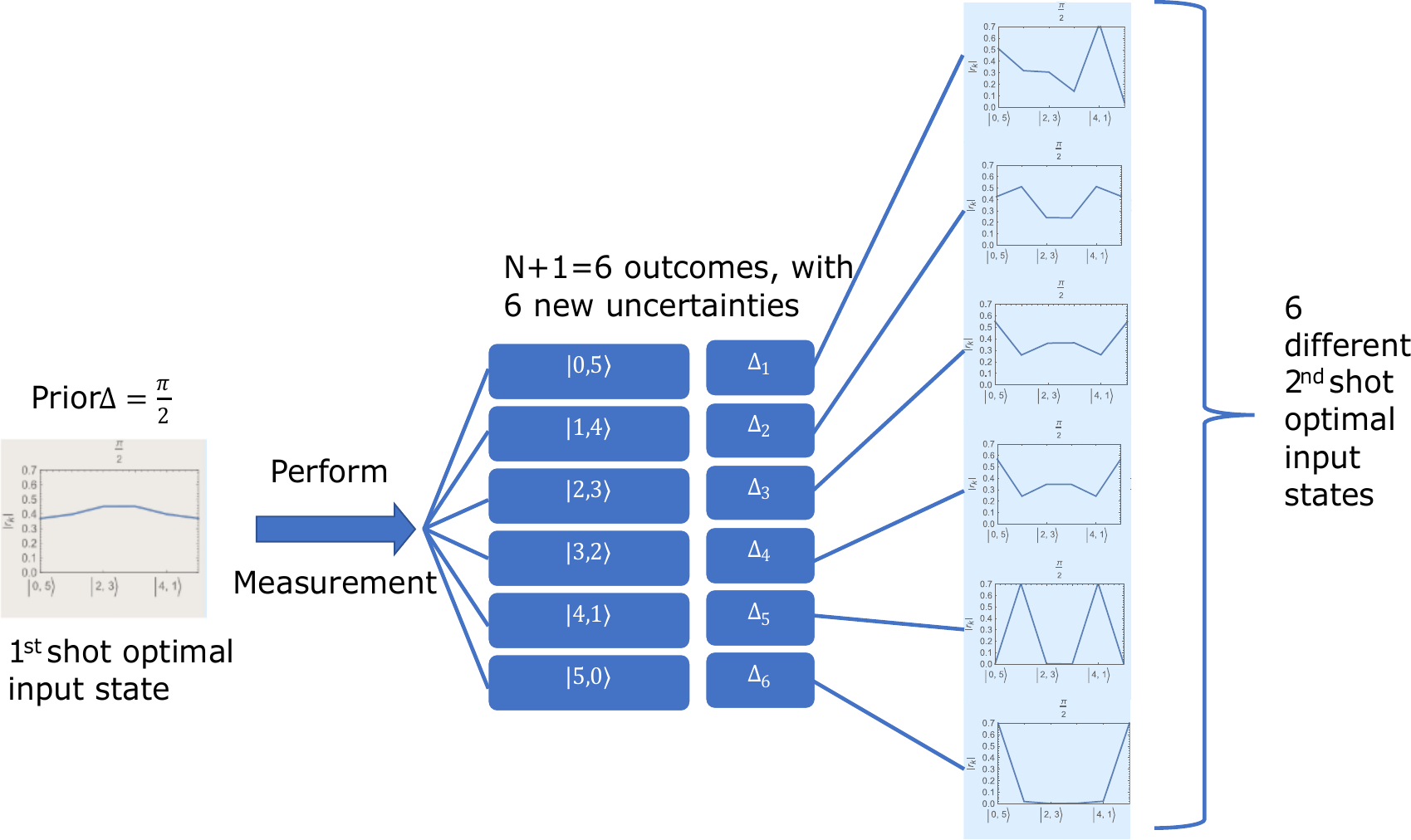}
    \caption{Optimal inputs for $N=5$ photons, $\nu=2$ shots obtained using global adaptive optimization for a specific value of the prior uncertainty, $\Delta=\frac{\pi}{2}$. These results correspond to the ones framed in orange in Fig.~\ref{Adaptive global optimal state}.}
    \label{Adaptive global optimal state2}
\end{figure*}

The optimal input states are obtained by minimizing the BMSE in \eqref{variance seq upd} w.r.t. the input state parameters $r_k$ and $\theta_k$. Fig.~\ref{Adaptive global optimal state} shows the optimal input state amplitudes $|r_k|$ for various prior uncertainties $\Delta=\pi,\frac{\pi}{2},\frac{3\pi}{10},\frac{\pi}{10}$, each plotted on a different row. In each row, the plot with a light brown background represents the optimal input state for the $1^{st}$ shot. The plots with a light blue background in each row are the optimal input states for each of the possible $N+1=6$ outcomes of the corresponding $1^{st}$ shot measurement. For example, the row for $\Delta=\frac{\pi}{2}$ (framed in orange) is shown more explicitly in Fig.~\ref{Adaptive global optimal state2}.

\subsection{Adaptive Local Optimization -- Feedforward}\label{Feedforward}
As in the non-adaptive case (Sec.~\ref{SBS}), we can choose to optimize the BMSE in \eqref{variance seq upd} locally (or shot-by-shot). Again, due to the complexity of the feedforward problem, we focus on the 2-shot example, as we did in Sec.~\ref{Adaptive Global}. This shot-by-shot method can be extended in a straightforward manner to many shots. For the 2-shot adaptive local optimization method, the protocol is as follows:
\begin{enumerate}
    \item Assume a flat prior probability distribution, of width $\Delta$, before the $1^{st}$ shot
    \item Make a measurement using the input state that is optimal for prior uncertainty $\Delta$
    \item Note down the phase estimator \eqref{phitilde seq upd} as well as the variance or BMSE \eqref{variance seq upd loc} after this measurement. Here we do not sum over the outcomes, instead keeping the individual posterior variance for each of the outcomes $m_1$ separate:
    \begin{multline}
    (\delta\phi)^2_{m_1}= \sum_{m_2}p(m_2|m_1) \times \\
     \frac{\int_{-\frac{\Delta}{2}}^{\frac{\Delta}{2}} (\phi - \tilde{\phi}_{m_2,m_1})^2 p(\phi) p(m_1|\phi)p(m_2|\phi,m_1)\, d\phi\ }{p(m_2,m_1)}  \,. \label{variance seq upd loc}
     \end{multline}
    \item We now have $N+1$ individual posterior variances $(\delta\phi)^2_{m_1}$ for $m_1=1 \ldots N+1$, one for each of the $N+1$ possible outcomes of the $1^{st}$ shot measurement
    \item Convert these $N+1$ individual variances into posterior phase uncertainties $\Delta_{m_1} = \sqrt{12 (\delta\phi)^2_{m_1}}$
     \item For each branch, the posterior uncertainty $\Delta_{m_1}$ after the $1^{st}$ shot becomes the new prior uncertainty for the $2^{nd}$ shot
    \item The optimal input state for the second shot is the one that is optimal for phase uncertainty $\Delta_{m_1}$, but shifted by $\tilde{\phi}_{m_1}$ to account for the fact that the new PDF is not centered on $\phi=0$
    \item We thus obtain a set of $N+1$ second shot optimal states -- one for each of the $N+1$ outcomes of the first shot
\end{enumerate}
 
\begin{figure}[h]
    \centering
    \includegraphics[width=0.7\columnwidth]{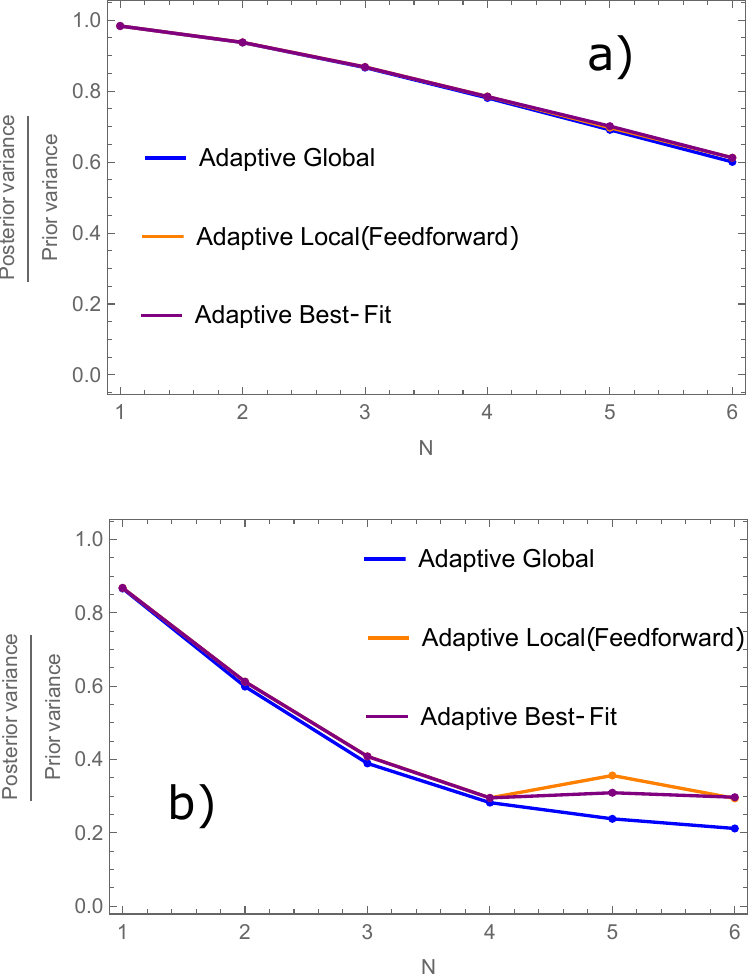}
    \caption{Variance ratio for a 2-shot sequence as a function of photon number $N$. Panels a) and b) show results for initial phase uncertainty $\Delta=\frac{\pi}{10}$ and $\Delta=\frac{3\pi}{10}$ respectively.}
    \label{fig:Adaptive method comparison}
\end{figure}

As we did before, instead of optimizing over the full space of coefficients $c_k=r_k e^{i \theta_k}$ in \eqref{inputstate} to find each optimal input state, we can use the best-fit formulae found in Sec.~\ref{Analytical formulae}. In Fig.~\ref{fig:Adaptive method comparison}, we compare three adaptive methods for finding the optimal input states:
the adaptive global optimization approach of Sec.~\ref{Adaptive Global}, the adaptive local optimization approach presented above, and the adaptive local approach simplified by using at each step the ``best-fit'' formulae of Sec.~\ref{Analytical formulae} instead of performing optimization.
Again, we see that the best-fit formulae provide excellent performance at trivial computational cost.

\begin{figure}[h]
    \centering
    \includegraphics[width=\columnwidth]{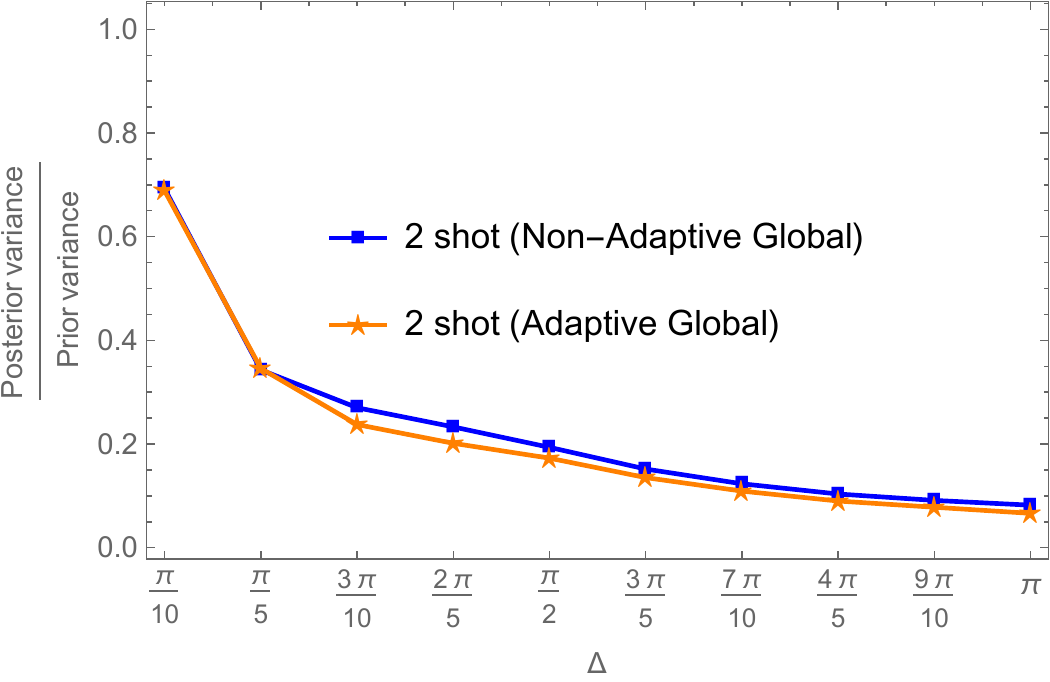}
    \caption{Global optimization strategy in the non-adaptive formalism of Sec.~\ref{Non-Adaptive formalism} \textcolor{blue}{(blue,square)} compared with the global optimization strategy in the adaptive formalism of Sec.~\ref{Adaptive formalism} \textcolor{orange}{(orange, star)}, as a function of the initial phase uncertainty $\Delta$, for $N=5$ photons and  $\nu=2$ shots.}
    \label{Comparing NA with A Global}
\end{figure}

Fig. ~\ref{Comparing NA with A Global} compares the adaptive and non-adaptive methods with global optimization for two shots, for different values of the initial phase uncertainty $\Delta$. We see that while the adaptive method gives a greater reduction in variance than the non-adaptive method, the improvement is only marginal.

Overall, the non-adaptive local strategy using the optimal N00N and Gaussian formulae of Sec.~\ref{Analytical formulae} appears to provide solid performance with the least resource demand of all the methods studied. 

\section{Monte Carlo Simulations}

The models in Sections~\ref{Non-Adaptive formalism} and \ref{Adaptive formalism} use Bayesian inference (a theorist's perspective) to determine the optimal phase estimation strategy. We will now compare these results to an analysis using frequentist inference (an experimentalist's perspective). We do this by performing Monte-Carlo simulations of the two shot-by-shot optimization methods discussed in Sections~\ref{SBS} and \ref{Feedforward}.

\subsection{Non-Adaptive Monte-Carlo Type Simulation (MCNA)}
Following the principles discussed in Sec.~\ref{SBS}, we start with a flat initial uncertainty interval $\Delta_{start} = \Delta_{in}$ such that the true phase $\phi_{true} \in [-\Delta_{start}/2,\Delta_{start}/2]$. Our initial phase estimator is $\tilde \phi_0=0$. Based on the discussion in Sec.~\ref{General}, if $N \Delta_{in} > 5$, we use a Gaussian input state for the first measurement; otherwise we use a N00N input. We then simulate a random measurement outcome $m_1$ using a random variable with probability distribution calculated by (\ref{eq: pmphi}). After this first-shot measurement, we obtain a new uncertainty interval and a shift $\phi_{new}$ in the phase estimator. The phase estimator after the first shot is now $\tilde\phi_1=\tilde\phi_0+\tilde\phi_{new}$. The process is repeated for $\nu$ shots. This is a local method, which is performed one shot at a time and therefore scales easily with $\nu$. As $\nu$ increases, we want to see $\tilde\phi_\nu$ (the phase estimator after $\nu$ shots) converge towards $\phi_{true}$.

\begin{figure}[h]
    \centering
    \includegraphics[width=\columnwidth]{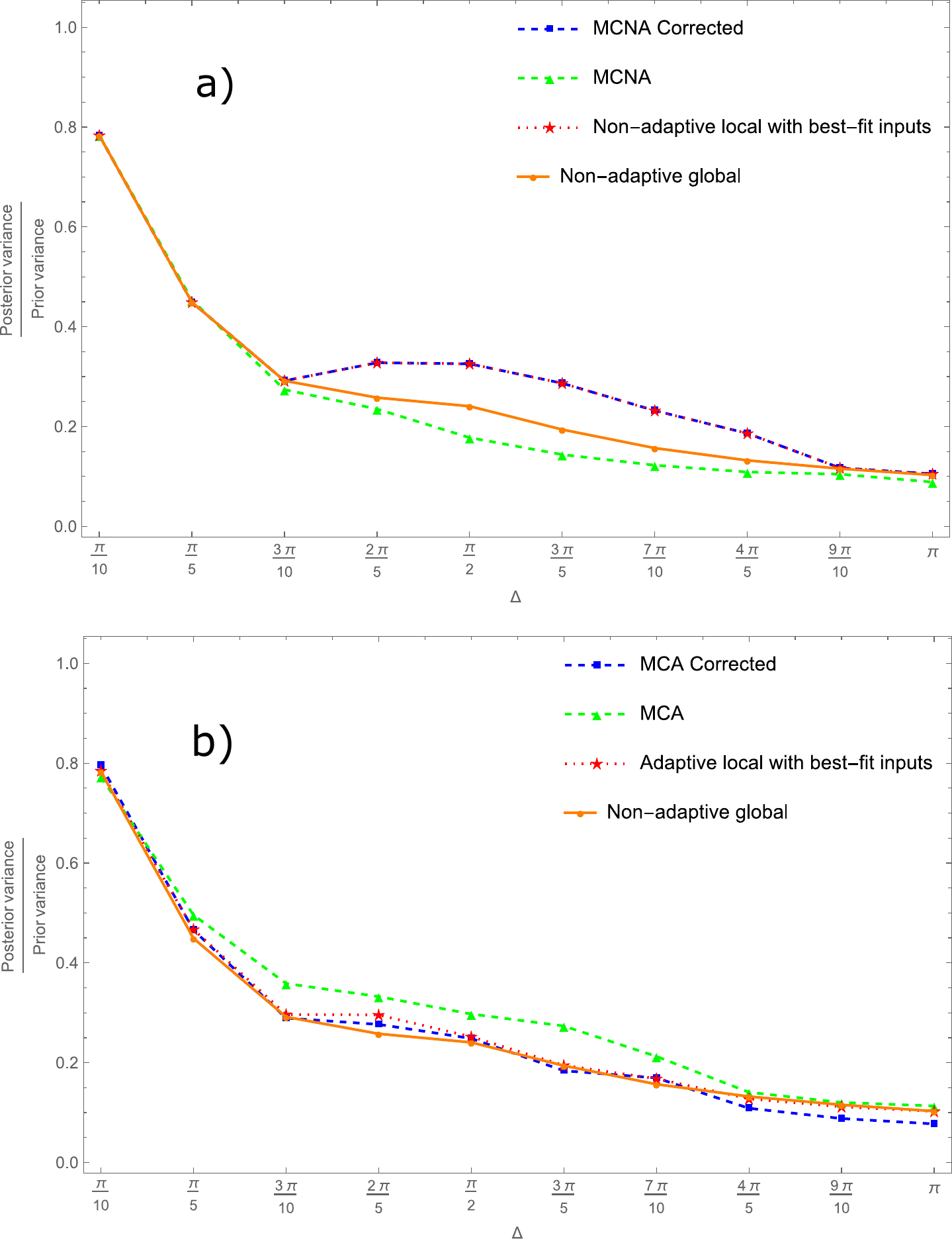} 
    \caption{Comparison of $\frac{\text{posterior variance}}{\text{prior variance}}$ obtained using Monte Carlo simulation with results from Sec.~\ref{SBS} in the non-adaptive case (Panel a) and with results from Sec.~\ref{Feedforward} in the adaptive case (Panel b). Here we consider the case of $N=4$ photons and $\nu=2$ shots.}
    \label{MCNA VarRatio Scaling delta}
\end{figure}

To compare the variance scaling of the Monte Carlo simulation with the results from Sec.~\ref{SBS}, we begin with $\nu=2$ shots. Averaging over many trials with $\phi_{true}$ chosen uniformly from $[-\Delta_{start}/2,\Delta_{start}/2]$ for each value of the initial phase uncertainty $\Delta_{start}$, we obtain the results labeled MCNA (Monte Carlo Non-Adaptive) \textcolor{green}{(green, triangle)} in Fig.~\ref{MCNA VarRatio Scaling delta}(a). 

Here it is important to note that this calculation is local (shot by shot), so in particular the prior distribution before the second shot is taken to be flat with width $\Delta'$, where $\Delta'^2/12$ is the variance after the first shot. The posterior variance after the second shot is then obtained using the flat distribution of width $\Delta'$ as the prior. Replacing the true posterior distribution after the first shot with a flat distribution of the same variance as the prior to the second shot introduces an approximation; the exact final variance given a set of two measurement results $m1,m2$ should instead be obtained by combining the two measurement results with the initial prior:
    \begin{equation}
    (\delta\phi)^2_{m_1,m2}= 
     \frac{\int_{-\frac{\Delta}{2}}^{\frac{\Delta}{2}} (\phi - \tilde{\phi}_{m_2,m_1})^2 p(\phi) p(m_1|\phi)p(m_2|\phi,m_1)\, d\phi\ }{p(m_2,m_1)}  \,, \label{variance one branch}
     \end{equation}
corresponding to one term in \eqref{variance seq upd}. Using \eqref{variance one branch} for each trial in the Monte Carlo simulation yields the MCNA Corrected \textcolor{blue}{(blue, square)} results in Fig.~\ref{MCNA VarRatio Scaling delta}(a), which as we see agrees perfectly with the Non-Adaptive local method \textcolor{red}{(red, star)} of Sec.~\ref{SBS}.

\begin{figure}[h]
\includegraphics[clip,width=0.8\columnwidth]{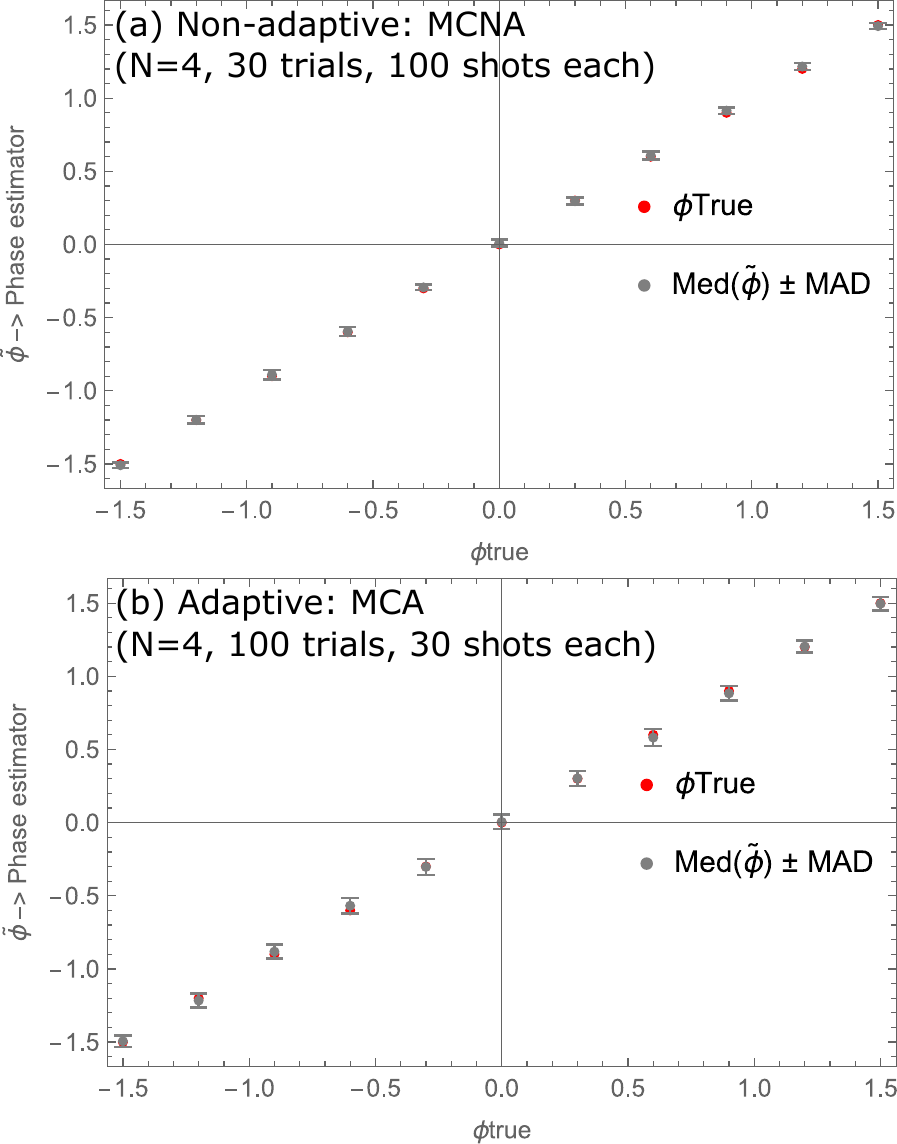}
\caption{The median value of the phase estimator $\tilde \phi$ over many Monte Carlo trials is calculated for each value of $\phi_{true}$ in the interval $[-\Delta_{start}/2,\Delta_{start}/2]$, where $\Delta_{start}=\pi$. The error bar in each case indicates the median absolute deviation (MAD) of $\tilde{\phi}$ from $\phi_{true}$.}
\label{medphitilde}
\end{figure}

In Fig.~\ref{medphitilde}(a), we extend the MCNA calculation to 100 shots and analyze the spread of the phase estimator values for each value of $\phi_{true}$ when the initial uncertainty is $\Delta_{start}=\pi$. In each case we show the median over 30 trials; the error bar indicates the median absolute deviation (MAD).

\begin{figure}[h]
    \centering
    \includegraphics[width=\columnwidth]{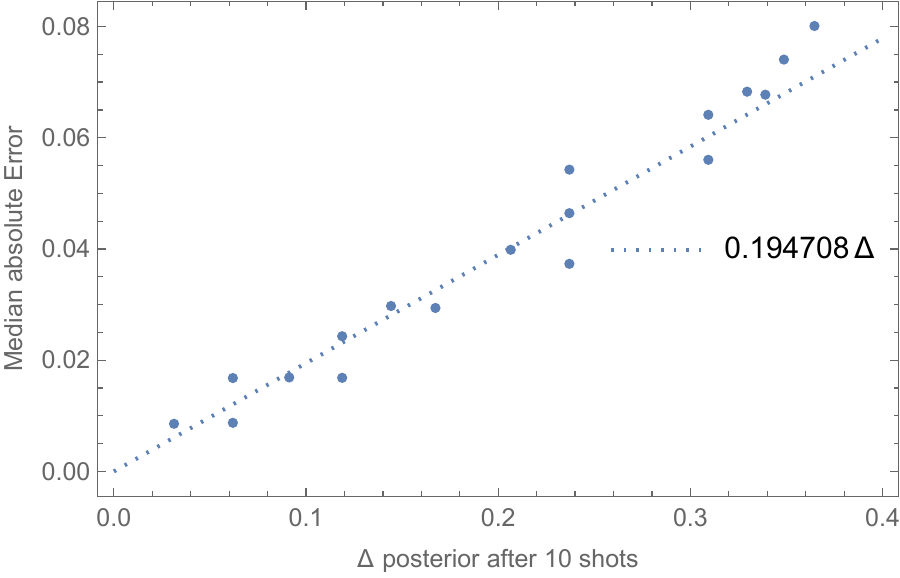}
    \caption{The average of the median absolute deviation (over different values of $\phi_{true}$) after 10 shots is proportional to $\Delta_{posterior}$. Here the number of photons is $N=4$.}
    \label{MAD propto deltaPosterior}
\end{figure}

We now take a given initial phase uncertainty $\Delta_{start}$ and average the MAD of the phase estimator after 10 shots over different values of $\phi_{true}$ in the interval $[-\Delta_{start}/2,\Delta_{start}/2]$. This is repeated for different values of $\Delta_{start}$ and in Fig.~\ref{MAD propto deltaPosterior} the resulting mean MAD after 10 shots is plotted versus the posterior uncertainty $\Delta_{posterior}$ after 10 shots. The result is compared with the predicted relationship between MAD and $\Delta_{posterior}$ for a Gaussian distribution of standard deviation $\sigma$: $MAD =  \sqrt{2} \erf^{-1}(1/2)\sigma= \sqrt{2} \erf^{-1}(1/2) \Delta_{posterior}/\sqrt{12}\simeq 0.1947 \Delta_{posterior}$. We confirm that the MAD is proportional to the posterior uncertainty $\Delta_{posterior}$, and that $\Delta_{posterior}$, which is computed in this ``experimental'' approach without reference to the true phase, is thus a good measure of the true error in the phase estimator. We also see that the results are reasonably consistent with the estimator being normally distributed around the true phase. 

\begin{figure}[h]
\includegraphics[clip,width=0.8\columnwidth]{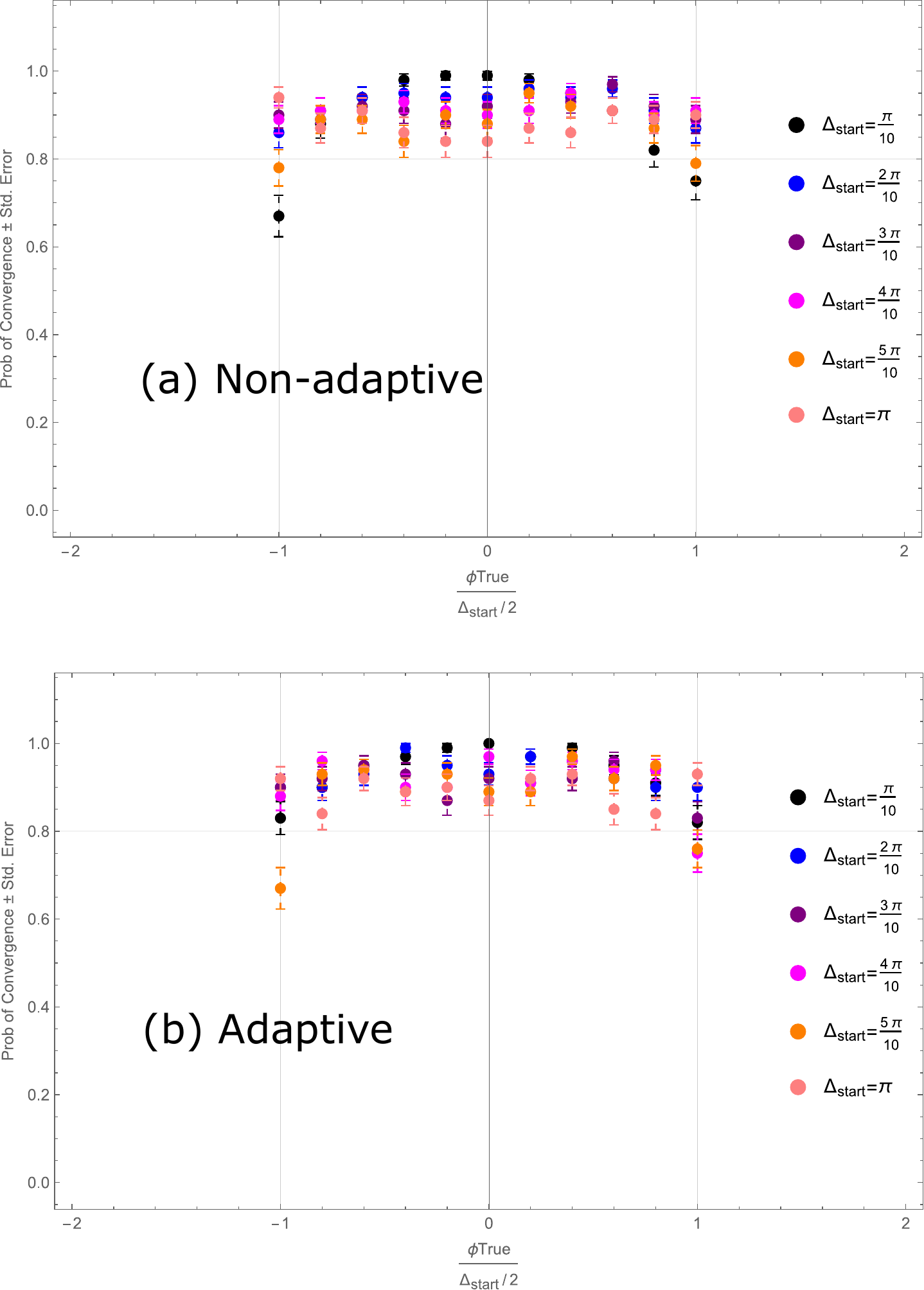}
\caption{The probability of the phase estimator $\tilde \phi$ after $\nu=10$ shots  being found inside the interval $[\phi_{true}-\Delta_{posterior}/2,\phi_{true}+\Delta_{posterior}/2]$, for different values of the initial uncertainty
$\Delta_{start}$ and for different values of the true phase $\phi_{true}$ within the initial uncertainty interval. Here the number of photons is $N=3$.}
\label{Probability of convergence_3}
\end{figure}

\begin{figure}[h]
\includegraphics[clip,width=0.8\columnwidth]{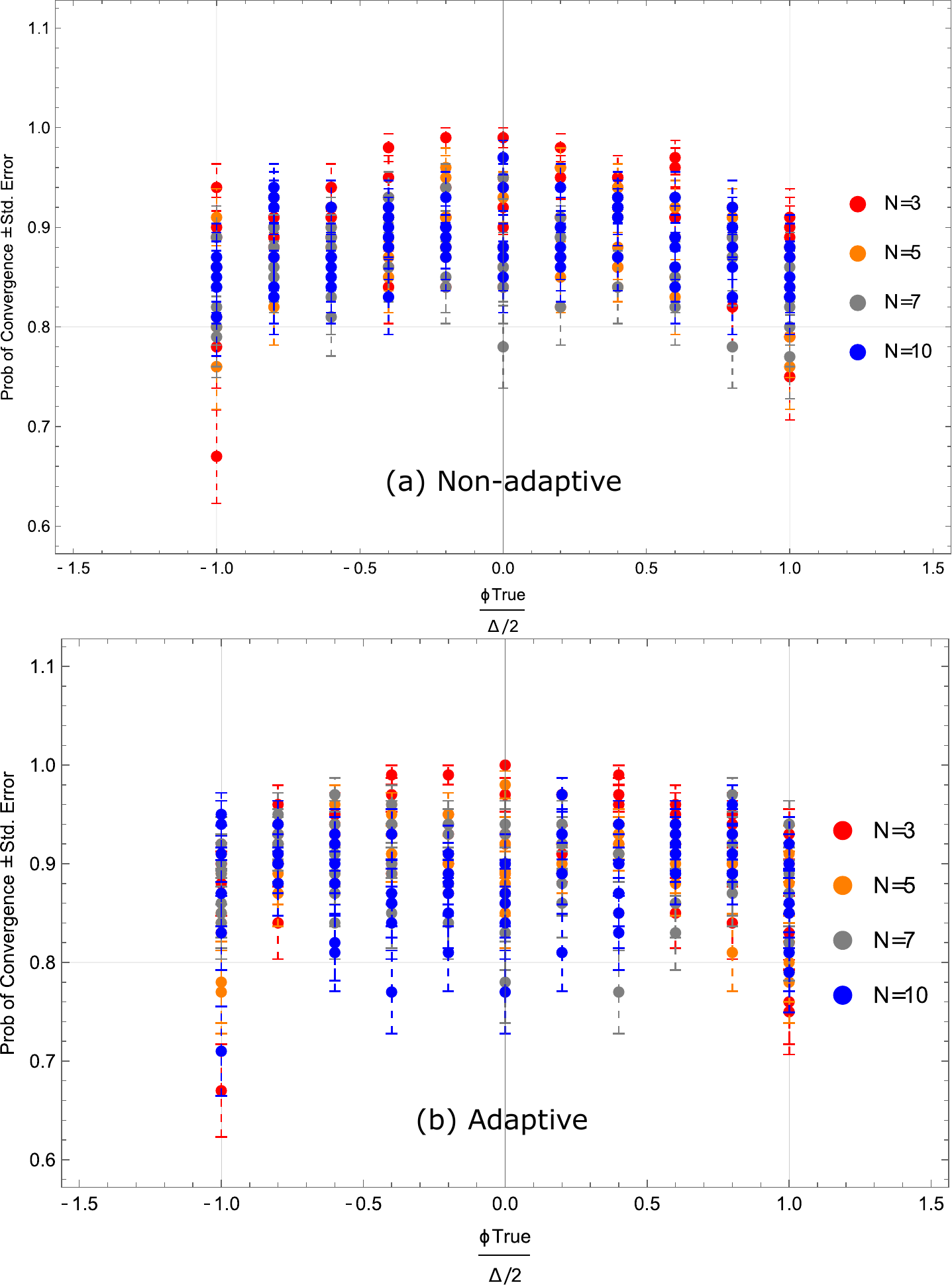}
\caption{The success probability after $\nu=10$ shots is computed as in Fig.~\ref{Probability of convergence_3}, but for different numbers of photons: $N$=3, 4, 7, and 10, and for the same six values of $\Delta_{start}$ shown in Fig.~\ref{Probability of convergence_3}.}
\label{Probability of convergence}
\end{figure}

We now turn to the tails of the distribution. In Fig.~\ref{Probability of convergence_3}, we plot the probability of the phase estimator $\tilde \phi$ after 10 shots with 3 photons being found within $\Delta_{posterior}/2$ of the true phase, i.e., inside the interval $[\phi_{true}-\Delta_{posterior}/2,\phi_{true}+\Delta_{posterior}/2]$. Since $\Delta_{posterior}=\sqrt{3} \sigma$, this ``success probability'' would be $\approx 91.7\%$ for normally distributed errors. We see from Fig.~\ref{Probability of convergence_3} that the actual success probability is lower in some cases, especially when the true phase is near the edge of the initial uncertainty interval ($\phi_{true}/\Delta_{start} \approx \pm 1/2$). Nevertheless, success probabilities >80\% are typically obtained even in the worst-case scenarios. Extending the analysis to different photon numbers $N$ in Fig.~\ref{Probability of convergence}, we again observe similar >80\% convergence probabilities.

\begin{figure}[h]
    \centering
    \includegraphics[width=\columnwidth]{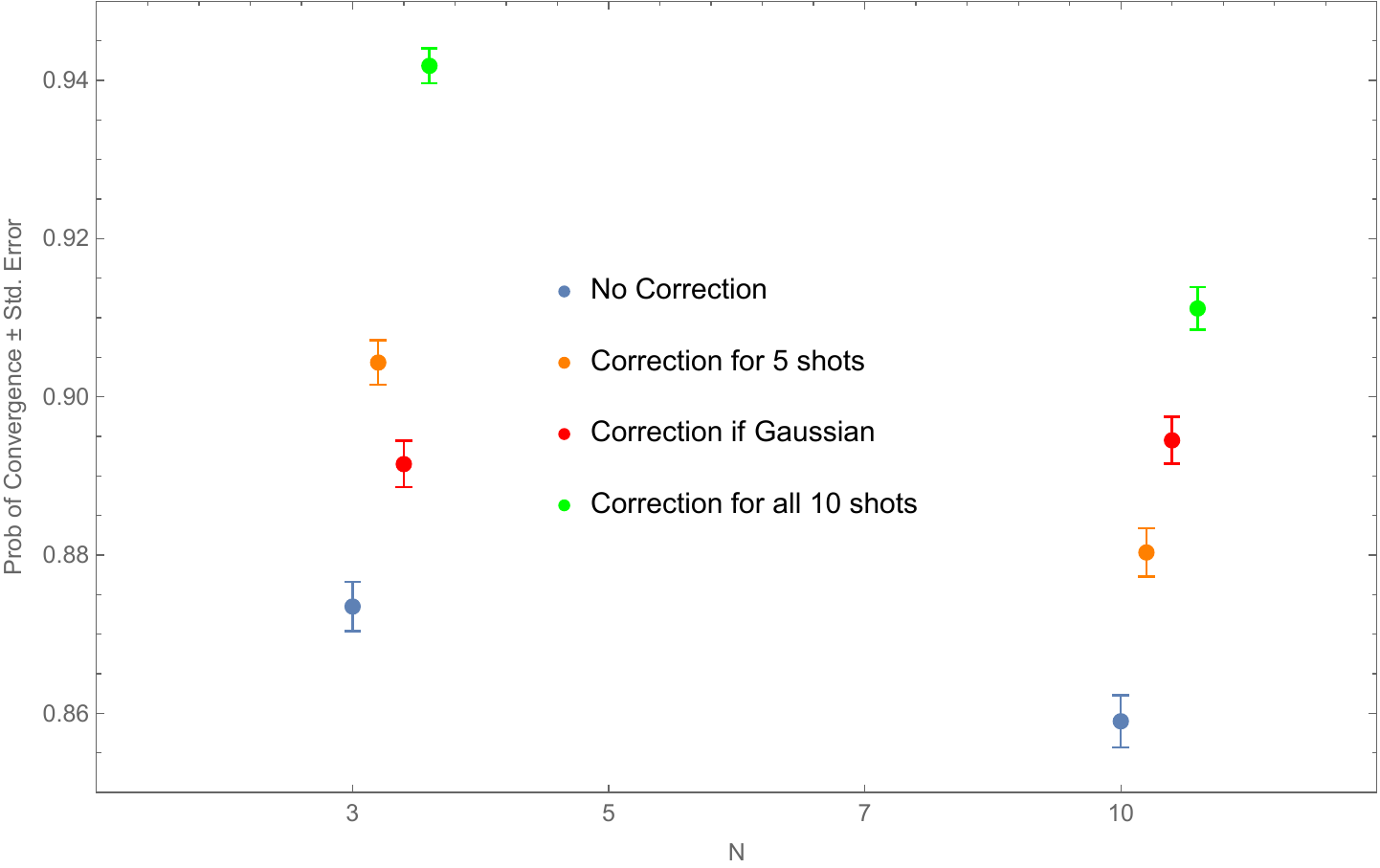}
    \caption{Several correction strategies are implemented to increase the MCNA success probability. Half of the optimal uncertainty reduction per shot is utilized \textcolor{orange}{for the first 5 shots (orange)}, \textcolor{red}{while the input state is Gaussian (red)}, or \textcolor{green}{for each of the 10 shots (green)}.}
    \label{CorrectionStrategyNA}
\end{figure}

Thus, in up to 20\% of trials, $\tilde\phi$ leaves the uncertainty interval sometime during the 10 shots and hence may not converge towards $\phi_{true}$. To address this, several possible correction strategies are implemented in Fig.~\ref{CorrectionStrategyNA} for $N=3$ and $N=10$. In each of these strategies, the uncertainty after a given shot is reduced by only half the amount prescribed by the optimal scaling formulae in \eqref{Gaussian scaling formula} and  \eqref{N00N scaling formula}. This more conservative approach is taken \textcolor{orange}{(i) for the first 5 shots only}, or \textcolor{red}{(ii) whenever we are in the Gaussian input state regime}, or \textcolor{green}{(iii) on all 10 shots}. Note that for the $N=10$ case in Fig.~\ref{CorrectionStrategyNA}, the uncertainty remains in the Gaussian regime during the first 7 shots, while for $N=3$ we stay in the Gaussian regime for the first 3 shots only.

While the correction strategies do provide a statistically significant improvement, they are likely not worth the effort, given that with an 80\% success probability it is cheaper simply to repeat the experiment in the case of failure.

\subsection{Adaptive Monte-Carlo Type Simulation (MCA)}
We perform a similar Monte Carlo analysis for the Adaptive strategy (MCA) as we did above for the Non-Adaptive case (MCNA). Results for MCA and MCA Corrected in Fig.~\ref{MCNA VarRatio Scaling delta} b). We observe that MCA is more prone to fluctuations and hence needs more trials to average over. For this reason, in Fig.~\ref{medphitilde}(b), we take 100 trials of 30 shots each instead of 30 trials of 100 shots. Finally, the convergence probabilities of MCA shown in Fig.~\ref{Probability of convergence_3}(b) and Fig.~\ref{Probability of convergence}(b) are very similar to those observed in the MCNA case, Fig.~\ref{Probability of convergence_3}(a) and Fig.~\ref{Probability of convergence}(a). Note that in some cases, e.g. $\Delta_{start}=5\pi/10$ and $\phi_{true}/\Delta_{start}=-1/2$ in Fig.~\ref{Probability of convergence_3}, the MCA (Adaptive) approach performs marginally {\it worse} than the MCNA (Non-Adaptive) approach. This is possible because the ``optimal'' protocol in each case is one that is optimal on average (averaging the unknown random variable $\phi_{true}$ over the uncertainty interval $[\Delta_{start}/2,\Delta_{start}/2]$) but of course this optimal state need not necessarily be optimal for every given value $\phi_{true}$. Here the anomalous behavior occurs in a situation where the true value is right at the edge of the prior uncertainty interval, and also we find ourselves in this example in the intermediate regime between Gaussian and N00N (see Fig.~\ref{Boundaries}).

\begin{figure}[h]
    \centering
    \includegraphics[width=\columnwidth]{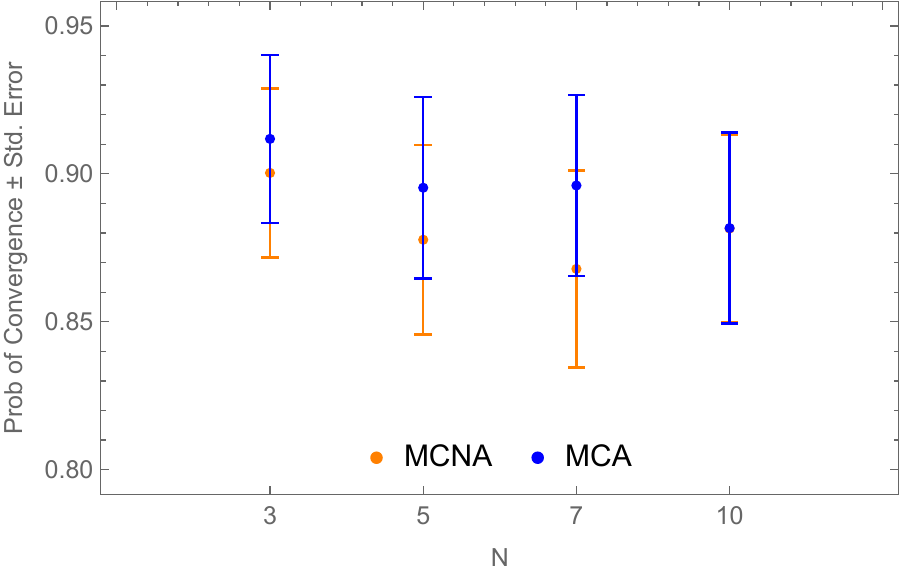}
    \caption{Comparing convergence probability of MCA with MCNA.}
    \label{Convergence probability NA_A}
\end{figure}

The Adaptive and Non-Adaptive strategies are compared directly in Fig.~\ref{Convergence probability NA_A}, where we average over $\phi_{true}$ and also average over all the initial uncertainties $\Delta_{start}$ that were considered in Fig.~\ref{Probability of convergence_3}. Fig.~\ref{Convergence probability NA_A} shows that the while the Adaptive approach performs somewhat better on average, the difference is not statistically significant.

\section{Conclusions}

We address the problem of optimizing Mach-Zehnder interferometer measurements with $N$ photons for a given initial phase uncertainty $\Delta$. Using Bayesian inference, we have constructed
the minimum mean squared error (MMSE) phase estimator, which 
minimizes the Bayesian mean square error (BMSE) or the posterior variance. We show that for a single measurement, N00N and Gaussian input states are excellent approximations to the optimal states in the $N\Delta<5$ and $N\Delta\geq 5$ regimes, respectively. These states are also close to optimal in the multi-shot scenario. The optimal N00N and Gaussian formulae were determined as functions of $N$ and $\Delta$ for a single shot, and the posterior variance obtained with these simple inputs almost perfectly matches the optimal posterior variance. This shows that the simple formulae can be used to construct the input states instead of performing a full optimization.

Next, a local non-adaptive (shot-by-shot) measurement strategy has been implemented. Here, N00N and Gaussian input states are again used, but after each shot the uncertainty $\Delta$ is adjusted and the optimal input state for the next shot is constructed using the updated valued of $\Delta$. Although this approach is not quite as good as  global optimization over all $\nu$ shots, it was shown that the results are comparable, while the scalability of the local approach is far superior to that of global optimization. This allowed the construction of general scaling formulae \eqref{Gaussian scaling formula} and \eqref{N00N scaling formula}  that show how many shots it will take on average to reduce phase uncertainty from a given initial value to a given target value. This number can be viewed as a benchmark that we can expect to match on average.

An adaptive measurement strategy was then implemented, where the input state for a given measurement is allowed to depend on the outcomes of the previous measurements. While the 2-shot adaptive method gave a greater reduction in variance than the 2-shot non-adaptive method, the improvement was only marginal.

Finally, we confirmed the 2-shot results found using Bayesian inference (a theorist's perspective) by performing a Monte-Carlo type simulation (an experimentalist's perspective), Extending the Monte Carlo approach to many shots, we confirmed that the phase estimator converges to the true value of the phase in most trials, with a Gaussian-distributed error. For $\nu=10$ shots, the probability of convergence was around 88\% for the non-adaptive simulation and around 90\% for the adaptive simulation.

In the present work, zero photon loss and perfect detector efficiency were assumed. The extension of the Bayesian approach to include finite loss in one of both arms of the interferometer, and inefficiency in the photon detectors, is left for future work.

\section*{Acknowledgments}

The authors thank Dmitry Uskov for helpful discussions.

\section*{Conflict of Interest}

The authors have no conflicts to disclose.

\section*{Author Contributions}

SS had the lead responsibility for performing the computations and writing the first draft of the manuscript. LK supervised the project. Both authors contributed equally to the theoretical development, data interpretation, and finalizing the manuscript.

\section*{Data Availability}

The data that support the findings of this study are available from the authors upon reasonable request.

\nocite{*}
\bibliography{library}

\end{document}